\documentclass[acmsmall,screen, preprint, nonacm]{acmart}
\usepackage{tabularx}
\usepackage{array}
\usepackage{wrapfig}
\usepackage{enumitem}
\usepackage{xspace}
\usepackage{xcolor}
\usepackage{enumitem}
\usepackage{adjustbox}
\usepackage[table]{xcolor}
\usepackage{pgfplots}
\pgfplotsset{compat=1.18}
\usepackage{listings}
\usepackage{needspace}
\definecolor{sigcell}{RGB}{232,241,250}
\definecolor{parttwo}{HTML}{A34F9E}
\definecolor{partthree}{HTML}{B76542}
\newcolumntype{Y}{>{\raggedright\arraybackslash}X}

\usepackage[most]{tcolorbox}
\usepackage{listings}

\newtcblisting{promptbox}[1]{
  enhanced,
  listing only,
  listing engine=listings,
  colback=black!2,
  colframe=black!55,
  colbacktitle=black!55,
  coltitle=white,
  fonttitle=\bfseries\small,
  title={#1},
  boxrule=0.7pt,
  arc=1.5mm,
  boxsep=0pt,
  left=7pt, right=7pt, top=7pt, bottom=7pt,
  toptitle=3pt, bottomtitle=3pt,
  before skip=3pt, after skip=3pt,
  listing options={
    basicstyle=\ttfamily\scriptsize,
    breaklines=true,
    breakatwhitespace=true,
    columns=fullflexible,
    keepspaces=true,
    showstringspaces=false,
    tabsize=2
  }
}
\AtBeginDocument{%
  }

\usepackage{twemojis}
\usepackage{xcolor}
\usepackage{colortbl}

\definecolor{sigcell}{HTML}{DCE9F5}

\definecolor{reacol}{HTML}{CA8A04}
\definecolor{deccol}{HTML}{7C3AED}
\definecolor{alcol}{HTML}{15803D}

\newcommand{\BothAligned}{\textit{FullyAligned}}  % decision and reasons match
\newcommand{\DecAligned}{\textit{DecAligned}}   % decision only
\newcommand{\ReaAligned}{\textit{ReaAligned}}     % reasons only  % reasons only
\begin{document}

% \title{Right Advice, Wrong Reasons: Effect of Reasoning Alignment on Trust and
% Reliance in AI Advisors}
\title{Right Advice, Wrong Reasons: How Decision and Reasoning Alignment Shape Reliance on AI Advisors}

\author{Eunkyu Park}
\correspondingauthor
\authornote{Work done at Carnegie Mellon University.}
\affiliation{%
  \institution{Carnegie Mellon University}
  \city{Pittsburgh}
  \state{PA}
  \country{USA}
}
\affiliation{%
  \institution{Seoul National University}
  \city{Seoul}
  \country{South Korea}
}
\email{eunkyup@andrew.cmu.edu}

\author{Vasudha Varadarajan}
\affiliation{%
  \institution{Carnegie Mellon University}
  \city{Pittsburgh}
  \state{PA}
  \country{USA}}
\email{vvaradar@andrew.cmu.edu}

\author{Wesley Hanwen Deng}
\affiliation{%
  \institution{Microsoft Research}
  \city{New York}
  \state{NY}
  \country{USA}
}
\email{wesleydeng@microsoft.com}

\author{Anisha Reddy}
\affiliation{%
 \institution{Carnegie Mellon University}
  \city{Pittsburgh}
  \state{PA}
  \country{USA}}
\email{amreddy@andrew.cmu.edu}

\author{Gunhee Kim}
\affiliation{%
  \institution{Seoul National University}
  \city{Seoul}
  \country{South Korea}}
\email{gunhee@snu.ac.kr}

\author{Motahhare Eslami}
\affiliation{%
  \institution{Carnegie Mellon University}
  \city{Pittsburgh}
  \state{PA}
  \country{USA}}
\email{meslami@andrew.cmu.edu}

\author{Jana Schaich Borg}
\affiliation{%
  \institution{Duke University}
  \city{Durham}
  \state{NC}
  \country{USA}}
\email{js524@duke.edu}

\author{Maarten Sap}
\correspondingauthor
\affiliation{%
  \institution{Carnegie Mellon University}
  \city{Pittsburgh}
  \state{PA}
  \country{USA}}
\email{msap2@andrew.cmu.edu}

\renewcommand{\shortauthors}{Eunkyu Park et al.}






\begin{abstract}
When people seek advice from AI on social situations, 
the system may recommend a decision they fully agree with but justify it with reasons they disagree with. 
Conversely, it may offer reasons they agree with, and yet reach a decision they reject. 
We distinguish \emph{decision alignment}, whether an advisor recommends the option favored by a user's measured values, from \emph{reason alignment}, whether its stated reasons align with the user's values. In this work, we ask how these two forms of alignment shape people's preferences for AI advisors and their willingness to use them in different roles. We conducted a within-subjects study ($N{=}121$) in which participants evaluated three AI advisors varying in decision and reason alignment across eight everyday moral dilemmas. Participants then chose which advisors they would consult, allow to draft messages for review, or allow to act on their behalf in new tasks. When seeking advice, participants most often preferred an advisor that reasoned through their prioritized values while recommending the opposing decision (40\%). 
However, when participants had to pick advisors for new tasks, their preferences shifted toward advisors that also recommended decisions aligned with their values, rather than those that only shared their reasons. Also, as participants considered tasks that gave AI advisors greater autonomy, they became more likely to opt out of using any advisor, from 12\% when consulting to 38\% when allowing an advisor to act on their behalf. Our findings suggest that decision and reason alignment shape advisor preferences differently across contexts of AI use, and that alignment with a user's values does not necessarily imply willingness to delegate.
\end{abstract}

\begin{teaserfigure}
  \includegraphics[width=\textwidth]{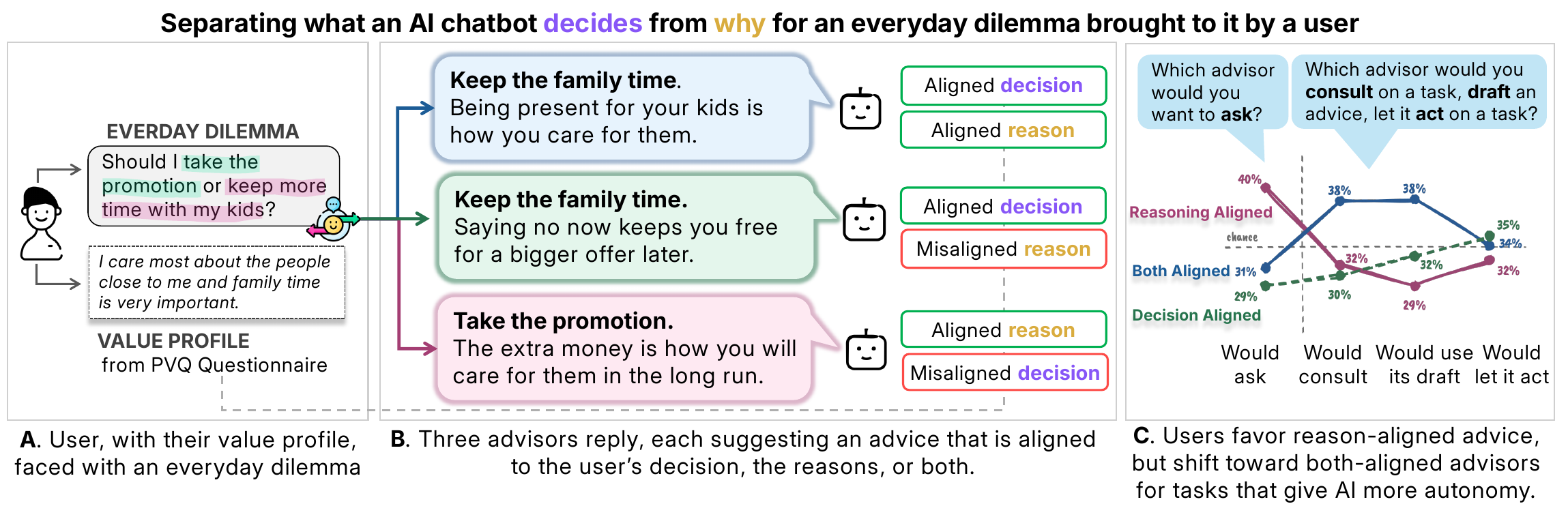}
   \caption{\textbf{Advisor preferences shift from evaluating advice to considering AI use.}
(A) A participant's value profile and an everyday dilemma.
(B) Their value scores determine the three AI advisor's reply on every dilemma, and advisor colors are counterbalanced across participants.
(C) Participants most often chose the reasoning-aligned advisor when selecting whom to ask for advice, but this preference weakened when they considered using the same advisors for new tasks.}
\Description{Three-panel overview of the study. Panel A shows a person facing a dilemma between accepting a promotion and spending time with family, along with their value profile. Panel B illustrates three AI advisors whose decisions algin with the person's values preferences, reasons align with the person's values or both. Panel C shows how advisor preferences in selection proportion shift from seeking advice to considering AI for tasks giving it more autonomy; advisors with aligned-reasons only were selected more when evaluating AI advices, but the fully aligned advisor gained preference in later tasks.}
  \label{fig:teaser}
\end{teaserfigure}

\maketitle

\section{Introduction}
People increasingly ask AI systems for advice about jobs, friendships, and relationships~\citep{chatterji2025chatgpt, tamkin2024clio}, and let them act on their behalf in tasks like drafting messages, screening job candidates, and moderating online forums~\citep{hohenstein2023ai, chan2023harms, wilson2024resume, kumar2024watch}. As AI systems take on these roles, an important question is whether their responses align with what the user values. Existing approaches to AI alignment often learn from human preference feedback or rule-based supervision~\citep{christiano2017deep, ouyang2022training, bai2022constitutional} but these objectives do not necessarily capture the values underlying those preferences --- a distinction that matters especially for moral advice, where values can differ across individuals and groups~\citep{kirk2024benefits, sorensen2024position}. Agreement on a recommended decision does not necessarily imply agreement on the values used to justify it, either: the same recommendation can be supported by different value-based reasons~\citep{chiu2026morebench}. This raises the question of whether matching a person's preferred decision is sufficient for alignment---that is, whether the AI advice is useful, reflects their underlying values and warrants their trust and reliance. 
In this work, we investigate how these dimensions of alignment can shape people's preferences for AI advisors, and whether these preferences persist when the AI takes on greater responsibility. We argue that alignment cannot be understood solely through agreement with a person's preferred decisions, and that its importance depends on the role an AI plays in their decision-making. 

To examine this, we distinguish an AI advisor's \emph{judgment}, the decision it recommends, from its \emph{reasoning}, the values it invokes to justify that decision.\footnote{We use \emph{judgment} for an advisor's verdict and \emph{choice} for a participant's selection among advisors. We use the term \emph{decision alignment} because an AI's judgment can become a decision made on someone's behalf.} Each dimension can independently align with or diverge from what the person values. We disentangle these dimensions---\emph{decision alignment} and \emph{reasoning alignment}---to investigate which people prioritize when choosing an AI advisor and whether their preferences change depending on the advisor's role.

Distinguishing reasoning from decision alignment matters because, while people assess AI advisors based on the relevant task competence as well as shared values~\citep{mayer1995integrative, siegrist2000salient, tolmeijer2022capable, mehrotra2021more}, they may agree with an advisor's recommendation while rejecting its underlying reasons, or vice versa. 
Neither agreement necessarily indicates that the advisor shares the user's values~\citep{chiu2026morebench}. 
Furthermore, acceptance of a model's judgment alone is an inconclusive signal of alignment: explanations can increase acceptance regardless of correctness~\citep{bansal2021does}, and agreement with an AI's judgment can reduce scrutiny of its reasoning~\citep{park2025critical} which could lead to overreliance on AI~\citep{kim2024m}.
Decisions and explanations therefore provide distinct, potentially conflicting cues about how an advisor might behave in future situations.
These cues may shape users' trust and reliance, which evolve with interaction experience, decision stakes, and perceived competence~\citep{li2023modeling, zhou2025rel}. 
Yet it remains unclear how reasoning and decision alignment actually influence people's willingness to use an AI advisor, particularly as its role shifts from providing advice to acting on their behalf. 

To investigate this, we develop a within-subjects user study where participants must choose between three advisors:
one aligned with the participant's value-favored decision but not their reasoning (\DecAligned{}), one aligned with their reasoning but not their decision (\ReaAligned{}), and one aligned with both (\BothAligned{}).
We present everyday dilemmas involving conflicting values grounded in Schwartz's theory of basic values~\citep{schwartz2012refining}. For example, a person who prioritizes family may face a choice between accepting a promotion and spending more time with their children (Figure~\ref{fig:teaser}). An advisor could recommend keeping family time because being present matters, recommend the same decision because it preserves future career opportunities, or recommend taking the promotion because the additional income could support the family. 
These different combinations of recommend decisions and value-based reasons justifying those decisions raise the question of which form of alignment people prefer in an AI advisor and whether those preferences persist when participants consider using the same advisors for related tasks, from consulting to drafting messages and acting on their behalf. We therefore investigate two main research questions:
\begin{enumerate}
    \item \textbf{RQ1:} When choosing which AI advisor to ask about a dilemma, do people favor one that shares their decision, their reasons, or both? (Section~\ref{sec:results-asking})
    \item \textbf{RQ2:} Does that preference hold when the advisor has a concrete role in a task, and does it change as that role gives the advisor more control? (Section~\ref{sec:results-shift})
\end{enumerate}

To answer these questions, we conducted a within-subjects study with 121 participants. We found that when choosing whom to ask for advice, participants preferred \ReaAligned{} (40\% of selections), choosing it more often than \DecAligned{} (+10.1\%, $p{=}.002$) or \BothAligned{} (+8.3\%, $p{=}.006$). However, this preference weakened when asked to consider the same advisors for new tasks: the within-participant contrast between \BothAligned{} and \ReaAligned{} shifted by 12.9\% toward \BothAligned{} ($p{=}.001$), although no advisor was consistently preferred within these tasks. Meanwhile, the proportion of choices selecting no advisor increased from 12\% for consultation to 38\% for autonomous action.

We end with a discussion of how alignment is defined, evaluated, and incorporated into AI systems that advise and act for people. Treating alignment solely as agreement with a preferred answer obscures whether an AI reaches the decision favored by a person's values or reasons from those values. Our findings show that these dimensions matter differently depending on the AI's role: reasoning alignment is particularly important when people seek advice, while neither reasoning nor decision alignment alone guarantees willingness to delegate. Evaluations of personalized AI systems should therefore distinguish decision alignment from reasoning alignment and account for the roles in which the systems will be used. For AI agents, aligning with a user's values may be necessary but is not sufficient to earn permission to act on their behalf. 

\section{Related Work}
\label{sec:related}

\subsection{Defining Alignment with a Person}

AI alignment concerns how systems can act in accordance with human instructions, intentions, and values~\citep{gabriel2020artificial}. Widely used approaches, including reinforcement learning from human feedback, learn from preferences expressed over model outputs~\citep{christiano2017deep, ouyang2022training}. 
However, even when an AI output matches a user's preferred answer, the reasoning behind the output may not be the same as the reasoning underlying the user's choice.
%response
%an AI shares the values underlying that preference. 
As preferences and values vary across individuals and groups, recent work has increasingly examined personalized and pluralistic alignment~\citep{kirk2024benefits, sorensen2024position, sepulveda2026people}. Moreover, the values expressed by an AI can shape both the positions it takes and the arguments it offers~\citep{liu2025synthetic}. From an HCI and FAccT perspective, alignment also unfolds through human--AI interaction~\citep{shen2025position, kim2024m, gao2026framing} and concerns both \emph{what} a system does and \emph{how} it does it~\citep{terry2023interactive, miller2023explainable}.
In personal and moral advice seeking scenarios, this distinction becomes especially relevant. Here, different values may support the same decision, and the same value may be used to justify competing decisions. These perspectives therefore motivate distinguishing an AI advisor's \emph{judgment}---the decision it recommends---from its \emph{reasoning}---the values invoked to justify that recommendation. Prior work shows that models can agree with users even when their stated views are incorrect~\citep{sharma2024towards, kim2024m}, and that agreement can shape users' evaluations of AI advice~\citep{cheng2026sycophantic}. While moral reasoning benchmarks have traditionally emphasized judgments, newer evaluations also examine the justifications behind them~\citep{jiang2025investigating, chiu2026morebench,  mun2025not, kingsley2024investigating}. Building on this work, we examine whether people prefer AI advisors that are aligned with their decisions, reasons or whether both matter.

\subsection{Trust and Reliance on AI Advice}

Trust involves a willingness to accept vulnerability to another party~\citep{mayer1995integrative}, whereas reliance concerns whether people act on that party's advice~\citep{lee2004trust, vereschak2021evaluate}. Research on appropriate reliance has examined how people decide whether to follow AI recommendations, including how confidence, uncertainty, and explanations shape their decisions~\citep{zhang2020effect, kim2024m, zhou2024relying, bansal2021does}. Reliance also depends on how people interact with AI systems~\citep{zhou2025rel}; notably, agreement with an AI's conclusion can reduce scrutiny of its reasoning~\citep{park2025critical}.
Understanding what shapes reliance is particularly important for personal and moral advice, where there may be no single objectively correct answer. Prior work shows that AI advice can influence moral judgments even when it is inconsistent~\citep{krugel2023chatgpt}, and that people may accept advice despite reservations about the system~\citep{bohm2023people, tolmeijer2022capable, landes2026people}. Beyond perceived competence, shared values can foster interpersonal trust~\citep{siegrist2000salient, twyman2008trust}, while perceived similarity between advisors and recipients can shape advice-taking~\citep{gino2009impact}. However, people may be less willing to rely on algorithms for subjective tasks~\citep{castelo2019task}. Whereas prior research has largely examined advice accuracy, explanations, and perceived trustworthiness, we investigate how people select among AI advisors whose recommended decisions and value-based reasoning align with them in different ways.

\subsection{From Asking to Delegating}
\label{sec:rw-stakes}

AI systems increasingly take on roles ranging from giving advice and helping with writing~\citep{chatterji2025chatgpt} to carrying out tasks with limited human involvement~\citep{handa2025economic}. These roles differ in the degree of control users retain~\citep{feng2025levels}. Recent work conceptualizes human--agent alignment as an ongoing interaction process shaped by how people specify their intentions, how agents decide when to act or defer, and how users evaluate and recalibrate their reliance on agents~\citep{long2026spear}. Even when people retain final approval, AI-generated suggestions can influence what they write and the opinions they express~\citep{jakesch2023co}. Whether users are willing to delegate also depends on the task, its perceived risks, and the control they retain~\citep{lubars2019ask, fugener2022cognitive}. People may be particularly reluctant to delegate morally consequential decisions~\citep{bigman2018people}, while delegation can also shift accountability and influence ethical behavior~\citep{steffel2016passing, kobis2025delegation, leib2024corrupted}. Thus, choosing an advisor to consult is distinct from authorizing it to act.

Prior work also suggests competing reasons why people might prefer different advisors across these forms of AI assistance. 
People often seek information that confirms their beliefs~\citep{hart2009feeling, kingsley2024investigating}, but may also seek advice to explore alternatives~\citep{bonaccio2006advice}. 
Research on human--AI complementarity suggests that altering perspectives and capabilities can support collaboration, while also creating tensions between agreement and complementary assistance~\citep{kumar2025seeing, amin2026align}. Conversely, agency theory emphasizes the importance of shared goals when a principal cannot fully monitor an agent~\citep{eisenhardt1989agency}. This suggests that alignment may become particularly relevant when users consider delegating actions. 
Together, these perspectives leave open whether people prioritize an advisor's recommended decisions or its underlying reasoning when considering different forms of AI assistance. Through an empirical study, we investigate this question by comparing participants' advisor selections when seeking advice and when considering hypothetical tasks involving consultation, drafting, and acting on their behalf.
\newcommand{\consult}{\textsc{Consult}\xspace}
\newcommand{\act}{\textsc{Act}\xspace}
\newcommand{\draft}{\textsc{Draft}\xspace}
\section{Study Design}
 
We ran a within-subjects online study in which each participant encountered three AI advisors, called AdvisorBot Blue, Green, and Pink, across eight everyday dilemmas.
All three advisors were built from the participant's own value profile, so that alignment is defined for each person going through the study (Figure~\ref{fig:dilemmas}).
One advisor matched both the decision the participant's values favor and the reasons behind it (\BothAligned), one matched only the decision (\DecAligned), and one matched only the reasons (\ReaAligned). The colored bots were randomly assigned for each user. 
Participants first observed the three advisors' responses to each dilemma and chose which advisor they would prefer to ask (RQ1). They then considered using the same advisors for new tasks, choosing which, if any, they would \consult{}, allow to \draft{} a message for review, or allow to \act{} on their behalf (RQ2). 
This allowed us to examine which dimension of alignment, the decision or reasons behind the advice, shaped initial advisor preferences, whether those preferences extended to new situations, and how advisor selection and willingness to delegate differed across roles involving varying levels of AI control. The study was approved by our institution's Institutional Review Board (IRB).

\subsection{User Study}
\subsubsection{Participants and their value profiles}
We recruited 121 participants from Prolific who lived in the US or the UK and spoke English as their first language.\footnote{Based on a power analysis of pilot data (n{=}72), detecting the primary within-person effect (reasoning- vs. decision-aligned advisor at the ask stage; pilot $d_z \approx 0.3$) with 80\% power at $\alpha = .05$ required roughly 75–100 participants, so we targeted around 120.} They were paid at \$12.55 per hour, and the median completion time was 21 minutes (Appendix~\ref{app:participants}). When they entered the study, all 121 participants completed the 21-item Portrait Values Questionnaire (PVQ-21)~\citep{schwartz2012refining}. Each item describes a person with particular values, and participants rated how similar that person was to themselves on a six-point scale.
\begin{figure*}[t]
  \centering
  \includegraphics[width=\textwidth]{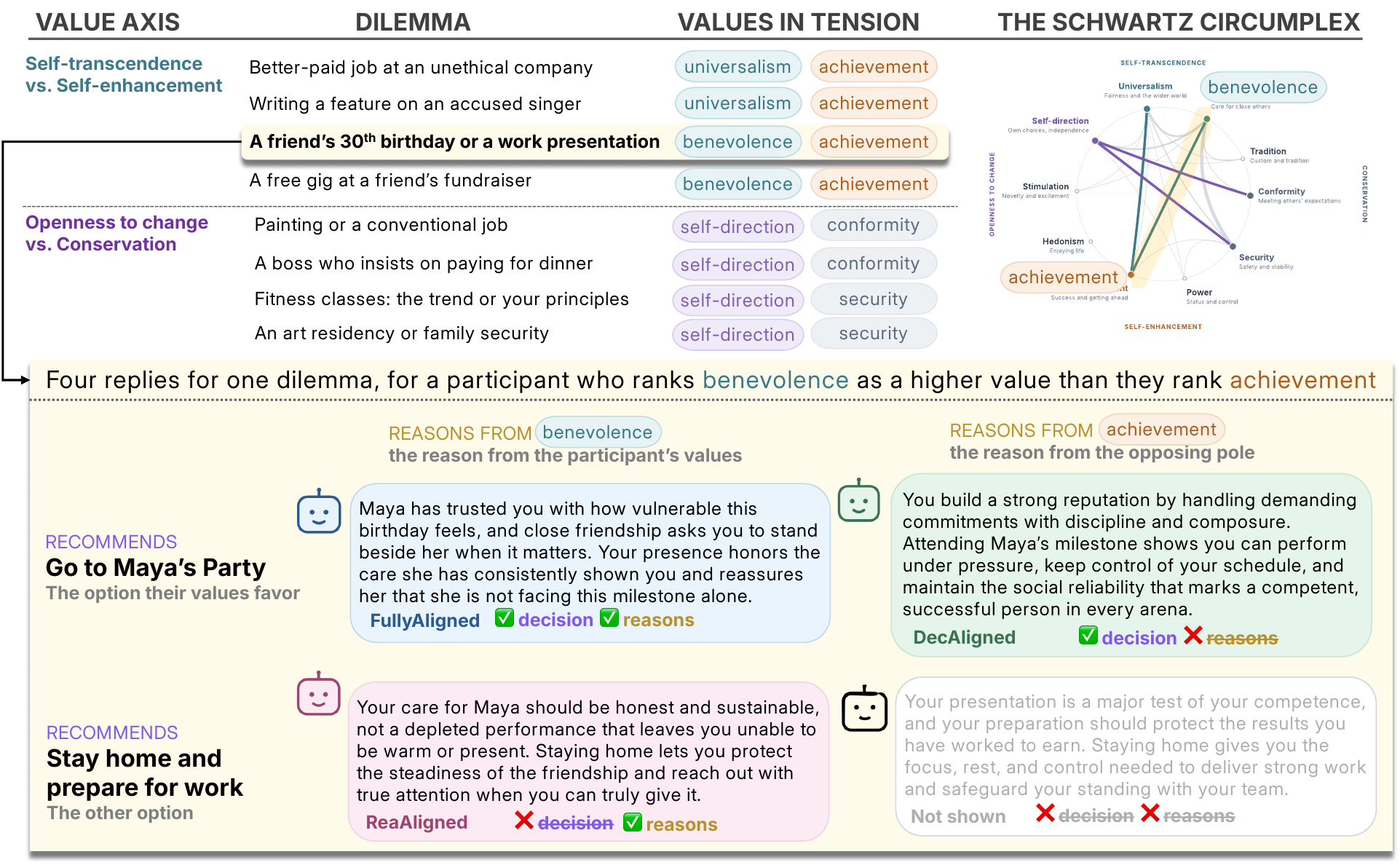}
  \caption{\textbf{Dilemmas and advisor replies.}
Top: Eight dilemmas, each involving a pair of competing Schwartz values. Bottom: An example crossing two recommendations with two value-based justifications. Advisor alignment conditions are assigned relative  to each participant's value profile; labels shown here assume benevolence is ranked above achievement. Only three of the four replies are presented to each participant. Full set of dilemmas and replies are shown in Appendix~\ref{app:dilemmas-full}
}
  \Description{Top: a table of eight everyday dilemmas grouped by value axis, four on self-transcendence versus self-enhancement and four on openness to change versus conservation. Each dilemma is shown with its two values in tension, next to Schwartz's value circle with a line joining each dilemma's two values in tension. Bottom: One dilemma, a friend's 30th birthday or a Monday presentation, is expanded below into a two-by-two grid of replies. Rows are the recommended option (go to Maya's party or stay home and prepare for work) and columns are the value the reasons come from, benevolence or achievement. Going for benevolence is FullyAligned, going for achievement is DecAligned, staying home for benevolence is ReaAligned, and staying home for achievement was not shown.}
  \label{fig:dilemmas}
\end{figure*}

\subsubsection{Dilemmas and replies}
\label{sec:dilemmas}
We constructed dilemmas and three different AdvisorBot replies that allowed us to vary the suggested decision and its reasons for the decision independently. 
We followed a five-step process to select dilemmas with conflicting values, rewrite them as natural requests for AI advice, and generate and validate replies (Appendix~\ref{app:validation}).

\textbf{Step 1: Pooling the dilemmas.} 
We started from the 200 everyday dilemmas in MoReBench~\citep{chiu2026morebench}, a benchmark of moral dilemmas posed to AI chatbots for advice, drawn from DailyDilemmas~\citep{chiu2025dailydilemmas}. Each has two options and an expert-written rubric naming the values in tension between them.

\textbf{Step 2: Mapping each tension on the Schwartz values.} GPT-5.5~\citep{openai2026gpt55} and Claude Opus~4.8~\citep{anthropic2026opus48} each mapped the tension onto a pair of Schwartz values.
We kept a dilemma only if (1) both models placed its two sides on opposite poles of the same axis, (2) named the same pair of values , (3) rated the fit high, and (4) judged the dilemma everyday and open to reasonable disagreement (filtering prompt in Appendix~\ref{app:tension-values}).
The models agreed on the value axis (1) for 172 of the 200 dilemmas (Cohen's $\kappa{=}.67$), and 18 passed every latter conditions.
We focused on dilemmas involving values at opposite poles of the same Schwartz axis, where each value clearly favors a different option. This allowed us to distinguish whether an advisor's recommendation or its underlying reasoning aligned with a participant's values.

\textbf{Step 3--4: Rewriting dilemmas and generating replies.} 
For each of the 18 dilemmas, GPT-5.5 rewrote the original third-person scenario as a first-person message ending in a question (e.g., ``...Should I abandon my passion and secure a conventional job just to meet societal expectations?''), to resemble how someone might naturally seek advice from an AI chatbot. We then generate four replies, crossing the two possible decisions with the two values in tension (Figure~\ref{fig:dilemmas}, bottom). A separate model, GPT-4.1-mini, checked whether each reply clearly expressed its intended decision and the reason, and whether replies suggesting the same decision offered distinct reasons.

\textbf{Step 5: Human Validation.}
Human annotators, who did not see the intended labels, identified which decision each reply recommended and which value its reasons for the decision resonated with. We required at least 85\% recovery of both the intended recommendation and value, and a difference of no more than 5\% in recovery accuracy between annotators who endorsed the reply's value and those who did not. The latter criterion was intended to ensure that replies were equally interpretable regardless of whether annotators personally shared the values expressed. In total, 42 annotators provided 408 judgments across 80 replies (approximately five judgments per reply, details in Appendix~\ref{app:building-human-val}).

\looseness=-1
\subsubsection{The stimuli for each participants}
\paragraph{The eight dilemmas.}
Eight dilemmas were selected as the final set of stimuli (Figure~\ref{fig:dilemmas}, top). Four involved self-transcendence versus self-enhancement: universalism versus achievement (e.g., taking a better-paid job at a company with questionable ethics) and benevolence versus achievement (e.g., attending a close friend's birthday or preparing for a work presentation). The other four involved openness to change versus conservation: self-direction versus conformity (e.g., paying for one's own dinner when a boss insists on paying) and self-direction versus security (e.g., leaving a stable job for an art residency). 
For each dilemma, the participant's relative scores on the two values determined which decision their values favored. 
Each re-written dilemma had a median length of 64 words and ended with a question. The two options were also presented as short imperatives (e.g., ``go to Maya's party'' and ``stay home and prepare for work''). The eight dilemmas appeared in randomized order. Full stimuli are provided in Appendix~\ref{app:dilemmas-full}.
\looseness=-1

\paragraph{Building three advisors for each participant.}
For each dilemma, we generated four replies by crossing the two possible decisions with the two competing values, such that each recommended decision could be justified by either value (Figure~\ref{fig:dilemmas}, bottom). Replies were addressed to ``you'' and argued for their option confidently, consistently grounding the argument in their value throughout the reply. To build the three advisors, we compared the participant's scores on the dilemma's two values for each dilemma per participant. 
We then used each participant's PVQ scores to determine which value they prioritized within the dilemma's value pair. 
We treated the option associated with their higher-scoring value as their \emph{value-favored decision}. We use the participants' measured value preferences to construct advisor alignment, rather than to predict their actual responses, as the same
value can motivate different actions, while the same
action can reflect multiple underlying values
~\citep{lee2022value}.  
This allowed us to construct three advisor conditions, each differing in whether its recommended decision and expressed reasons aligned with the participant's values:
\begin{itemize}
    \item \textbf{Fully-Aligned advisor (\BothAligned{})} that recommends their decision, with reasons from their value.
    \item \textbf{Decision-Aligned advisor (\DecAligned{})} that recommends their decision, with reasons from the other value.
    \item \textbf{Reason-Aligned advisor (\ReaAligned{})} that recommends the other option, with reasons from their value.
\end{itemize}

Each advisor maintained the same alignment condition for a given participant across all eight dilemmas. 
For example, consider a participant who scores higher on benevolence than achievement in the dilemma of attending Maya's birthday party versus preparing for a work presentation (Figure~\ref{fig:dilemmas}, bottom). \BothAligned{} recommends attending the party because being present demonstrates care for Maya. \DecAligned{} also recommends attending, but emphasizes how doing so demonstrates competence and strengthens one's reputation. \ReaAligned{} instead recommends staying home, arguing that caring for Maya also means sustaining the friendship without overextending oneself. 
Thus, \BothAligned{} and \DecAligned{} recommend the same option for different reasons, while \BothAligned{} and \ReaAligned{} uses the same value to justify different decisions. 
% Our goal is therefore to systematically distinguish alignment in what an advisor recommends from alignment in the values it uses to reason for that decision.

\subsection{Procedure}
\label{sec:procedure}
\begin{figure*}[t]
  \centering
  \includegraphics[width=\textwidth]{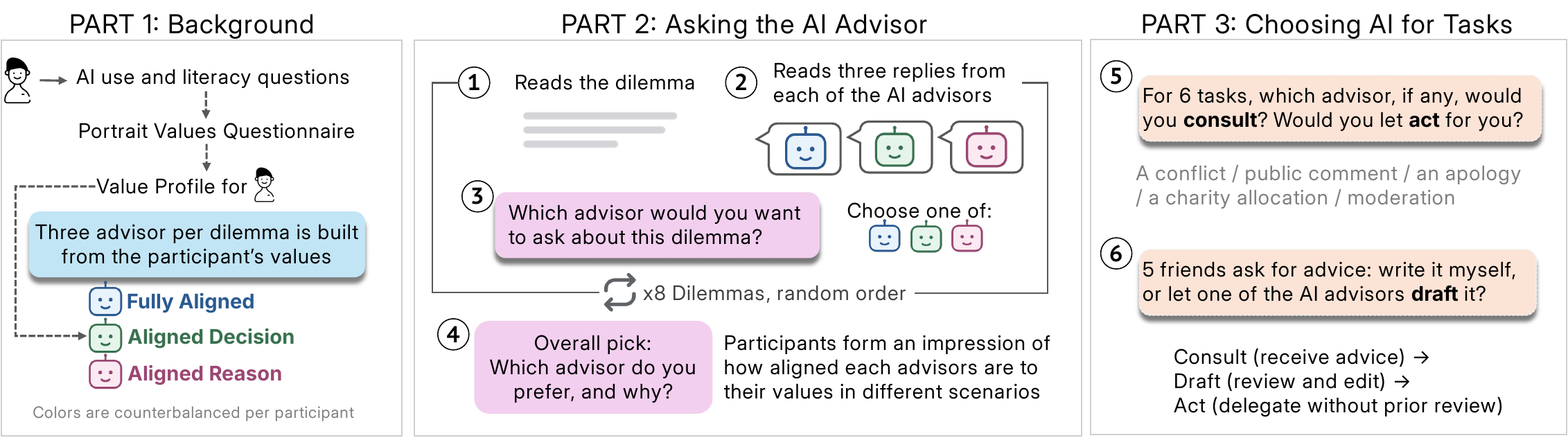}
  \caption{\textbf{Study procedure.}
  (1)~Participants report their AI use and AI literacy and complete the PVQ-21. Their value scores determine the three advisors' alignment condition. Advisor colors are counterbalanced.
  (2)~Across eight dilemmas, participants compare the three advisors' replies and choose whom they would want to ask (\textcolor{parttwo}{RQ1}). 
  % After the dilemmas, all participants rate each advisor.
  (3)~For six tasks, participants choose which advisor, if any, they would \consult{} and let \act{} for them. For five friends' messages, they choose whether to write the reply themselves or let an advisor \draft{} it (\textcolor{partthree}{RQ2}). A recap of each advisor's decisions is available throughout Part~3.}
  % Table~\ref{tab:questions} gives the wording of each choice, and Appendix~\ref{app:interface} shows the screens.}
  \Description{A three-part flow for the study, in the order of left to right. Part 1 collects answers to AI use and literacy questions and the Portrait Values Questionnaire, from which three advisors are built for each dilemma: FullyAligned, DecAligned, ReaAligned. Next, in part 2 participants read the dilemma, read three replies from the advisors, and choose which advisor they would want to ask, eight times in random order; then they make an overall pick with a reason and rate each advisor. Part 3 asks the participants to select one of the advisors, or none, for consulting them for a task, letting it act on their behalf, and drafting an advice for a friend.}
  \label{fig:procedure}
\end{figure*}
The study had three parts (Figure~\ref{fig:procedure}), with screenshots of the interface provided in Appendix~\ref{app:interface-capture}.

\textbf{Part 1: Background.}
Participants answered questions about how often they use AI chatbots and questions on AI literacy (details in Appendix~\ref{app:literacy}), then completed the Portrait Values Questionnaire.

\textbf{Part 2: Asking the AI advisor.}
This part served two purposes. First was to measure which advisor participants would prefer to ask if they faced a similar dilemma themselves (RQ1), and to familiarize them with how each advisor made decisions and reasoned for them before choosing advisors for different tasks in Part~3. 
For each of the eight dilemmas, participants read the situation as a chat message and saw the three advisors' replies side by side, with their display order randomized across dilemmas. 
They then chose which advisor they would want to ask if they were to get advice on making a similar decision themselves. 
We randomly selected four dilemmas to included attention checks about the situation's content.
After completing all eight dilemmas, participants 
selected their overall preferred advisor and explained their choice in an open-text response. These repeated exposures were intended to help participants form impressions of each advisor's decision and reasoning tendencies before deciding whether to use them in Part~3.

\textbf{Part 3: Using the advisor for concrete tasks.}
Part~3 examined whether the participants' advisor preferences carried over to tasks with different levels of AI involvement. 
Motivated by prior work on human-AI reliance~\citep{zhou2025rel}, we examined three roles: consulting an advisor (\consult), allowing it to draft a message for review (\draft), and allowing it to act on the participant's behalf (\act). These differed in how much control participants retained over the final action.

We designed six tasks that were varied along two dimensions that may shape willingness to delegate to AI: whether the outcome bears on the participant's identity, and whether it affects other people. Prior work suggests that identity-related concerns can shape acceptance of AI autonomy~\citep{choudhuri2026you}, while people are reluctant to delegate morally consequential decisions affecting third parties to machines~\citep{gogoll2018rage}. 
The tasks (Appendix Table~\ref{tab:part3-tasks}) covered interpersonal communication (a conflict with someone close, a provocative public comment, and an apology), resource allocation (a charity donation), and decisions affecting others (screening job candidates and moderating content). For each task, participants chose which advisor, if any, they would consult on how to handle the situation, and which, if any, they would allow to act on their behalf. Both questions included a ``None of them'' option. 

For the drafting role, participants read messages from five friends seeking advice. Each message was adapted from one of the earlier dilemmas, preserving its pair of opposing higher-order value axes and the type of relationship involved while changing the specific situation. Participants chose whether to write a response themselves or let one of the three advisors draft it, knowing they could review and edit the draft before sending it. Throughout Part~3, participants could access a recap of the advisors' earlier responses and their own choices from Part~2. Unlike Part~2, participants selected advisors for new tasks without seeing how they would respond. We therefore allowed participants to decline AI assistance, so that their choices could reflect not only which advisor they preferred, but also whether they considered any advisor suitable for the task based on their earlier impressions. Full stimuli and questions for both Part~2 and Part~3 are in Appendix~\ref{app:stimuli-wording}.
% \begin{table*}[t]
%     \centering
%     \small
%     \caption{Comparison of advisor selection across study parts. In Part~2, participants selected advisors after observing their responses to value-based dilemmas. In Part~3, participants selected advisors for new tasks based on impressions formed earlier, without seeing how the advisors would respond.}
%     \label{tab:study-comparison}
%     \begin{tabularx}{\textwidth}{@{}lXXX@{}}
%         \toprule
%         & \textbf{Part 2: Ask}
%         & \textbf{Part 3: Consult / Act}
%         & \textbf{Part 3: Draft} \\
%         \midrule
%         \textbf{Situations}
%         & 8 everyday dilemmas
%         & 6 new everyday tasks
%         & 5 new situations involving friends \\
%         \textbf{Advisor responses}
%         & All three replies shown
%         & No new replies shown
%         & No new replies shown \\
%         \textbf{Participant choice}
%         & Which advisor to ask
%         & Which advisor to consult or let act
%         & Write themselves or let an advisor draft \\
%         \textbf{Opt-out option}
%         & No
%         & Yes (``None of them'')
%         & Yes (``Write it myself'') \\
%         \textbf{Details}
%         & Establishes impressions of each advisor
%         & New tasks varying in personal and third-party stakes
%         & Preserves competing values and relationship types from Part 2 \\
%         \bottomrule
%     \end{tabularx}
% \end{table*}
\looseness=-1
\subsection{Analysis Methods}
\label{sec:analysis-approach}

We analyzed the advisor preferences separately for asking (RQ1) and for tasks with more AI involvement (RQ2). 
For each participant and study step, we calculated the \emph{proportion} of advisor selections going to each of the three alignment conditions, averaging these proportions across participants. 
To compare changes in advisor preferences between Parts~2 and 3, we calculated selection proportions conditional on choosing an advisor, excluding ``None of them.'' This accounts for the difference in choice sets: participants were required to select an advisor in Part~2 but could opt out in Part~3. We separately analyzed opt-out proportions in Part~3 using all choices as the denominator.  Also, we compared alignment conditions using within-participant differences in advisor selection proportions in \%. We used two-sided sign-flip permutation tests to assess differences and obtained 95\% confidence intervals through participant-level bootstrap resampling (10,000 resamples). To examine how advisor preferences changed when participants considered using the same advisors for new tasks (\consult{}, \draft{}, and \act{}), we compared each participant's pairwise selection differences in Part~2 with their corresponding differences aggregated across Part~3.
Finally, we conducted exploratory analyses of whether preferences varied with participants' value strength, defined as the mean absolute difference between the two competing value scores across the eight dilemmas. We also examined participants' open-text explanations of their overall advisor preferences.

\section{Results}
\label{sec:results}
\begin{figure*}[t]
  \centering
  \includegraphics[width=\textwidth]{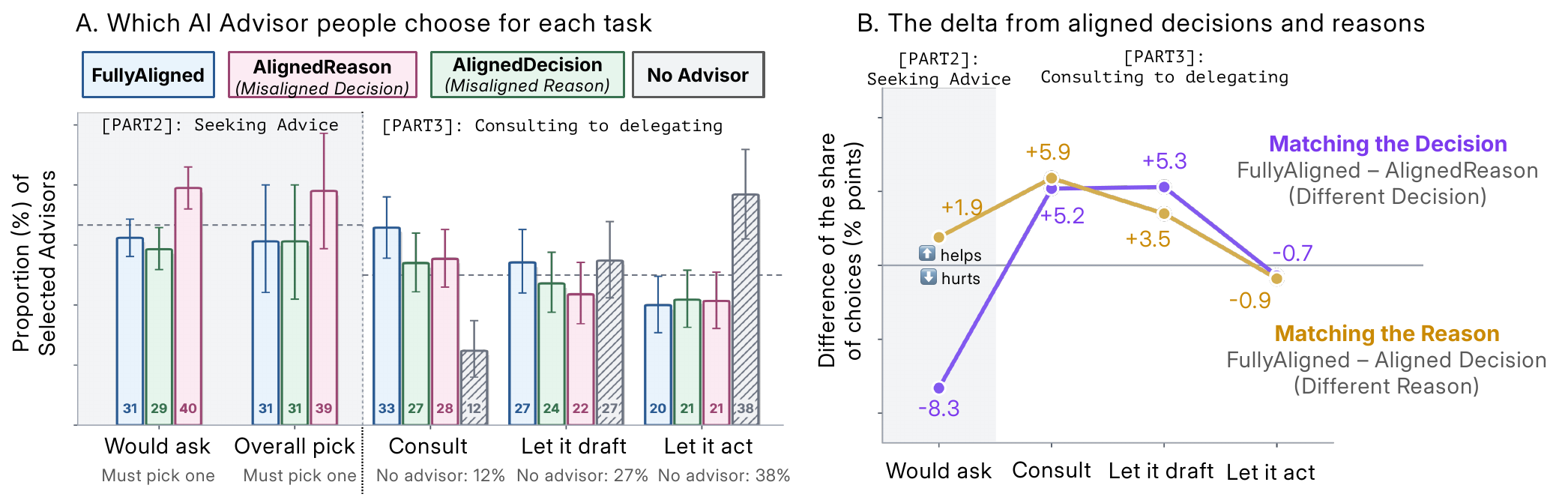}
  \caption{\textbf{Advisor selection across study settings.} 
  (A)~The proportion of each participant's advisor choices in Part~2 with 95\% CIs. (B)~Within-person gaps between Part~2 and Part~3 between two advisors that differ in the decision or reason. The numbers above zero means alignment in decision or reason increased preference (purple: \BothAligned{}$-$\ReaAligned{}) or decision (gold: \BothAligned{}$-$\DecAligned{}). }
  % Positive values favor \BothAligned{}.}
  \Description{Panel A compares selection proportion for FullyAligned, ReaAligned, DecAligned, and NoAdvisor across Part 2 and 3. Would ask: 31, 29, 40 percent. Overall pick: 31, 31, 39. Consult: 38, 30, 32, with 12 percent choosing no advisor. Let it draft: 38, 32, 29, with 27 percent choosing no advisor. Let it act: 34, 35, 32, with 38 percent choosing no advisor. A dashed line marks chance at 33 percent. Panel B plots the pairwise selection differences across Part 2 with two lines across would ask, consult, let it draft, and let it act. The purple line, matching the decision, goes from minus 8.3 to plus 6.7, plus 8.9, and plus 1.6. The gold line, matching the reasons, goes from plus 1.9 to plus 7.7, plus 5.9, and minus 1.0.}
  \label{fig:picks}
\end{figure*}

We first examine which advisor participants chose when asking for advice (RQ1) and how their preferences changed when they are asked to consider the use of AI advisors in specific tasks (RQ2). 
We also analyze how these results are different for participants with different value strengths and conclude with open-text feedback that shows how the different advisors were perceived by the participants. 
Unless otherwise noted, results are reported as proportions of advisor types selected by participants and within-participant differences in \%, following Section~\ref{sec:analysis-approach}. 

\subsection{When Asking, People Prefer the Advisor That Shares Their Reasons but Not Their Decision (RQ1)}
\label{sec:results-asking}
When deciding which advisor they would ask about a dilemma, participants selected \ReaAligned{} in 40\% of cases, compared with 31\% for \BothAligned{} and 29\% for \DecAligned{} (Figure~\ref{fig:picks}A). \ReaAligned{} was selected more often than \DecAligned{} by 10.1\% (95\% CI [3.9, 16.2], $p {=}.002$) and more often than \BothAligned{} by 8.3\% (95\% CI [2.7, 13.7], $p{=}.006$; Table~\ref{tab:contrasts}). \BothAligned{} and \DecAligned{} did not differ significantly (1.9 pp, $p{=}.54$). Surprisingly, pairwise comparisons between the two advisors that used reasoning grounded in the participant's values (\BothAligned{} vs.\ \ReaAligned{}) show that participants more often selected the advisor recommending the decision their values did \emph{not} favor. In contrast, when the advisors recommended the decision their value favored (\BothAligned{} vs.\ \DecAligned{}), reasoning alignment did not lead to a detectable difference in advisor preferences. 

These comparisons suggests that participant's preference for \ReaAligned{}'s may reflect the preference for an AI advisor presenting an alternative to their decision-aligned option while justifying it with reasons for the values they endorse, rather than a just a general preference for matching reasons alone. We explore possible explanations for this preference in Section~\ref{dis:alignment-deliberation}.

\looseness=-1
\subsection{Advisor Preferences and Willingness to Use AI at All Shift Across Task Contexts and Roles (RQ2)}
\label{sec:results-shift}
We next examined whether the preferences formed from observing advisors' responses in Part~2 carried over to selecting advisors for new tasks in Part~3. 
We inspect two aspects of these choices. First, how relative preferences among the three advisors changed across the two parts (Section~\ref{sec:Rq2-1}). Second, whether participants chose to use any advisor across the \consult, \draft, and \act roles (Section~\ref{sec:Rq2-2}). 
Together, in this section, we examine both which advisor participants preferred and their willingness to use AI assistance at all.
\begin{table*}[t]
    \centering
    \small
    \caption{\textbf{Pairwise differences in advisor selection proportions (percentage points).}
    Positive values favor the first advisor. Proportions exclude ``None of them'';
    $n$ denotes participants selecting at least one advisor.
    The change row compares \emph{Would ask} with Part~3 overall within participants.
    Brackets indicate 95\% bootstrap CIs.
    The final row reports advisor preference counts with exact binomial tests.}
    \label{tab:contrasts}

    \renewcommand{\arraystretch}{1.12}
    \begin{tabularx}{0.97\textwidth}{
        @{}lr
        >{\raggedright\arraybackslash}X
        >{\raggedright\arraybackslash}X
        >{\raggedright\arraybackslash}X@{}
    }
      \toprule
      Step & $n$
      & \shortstack[l]{Full vs.\ Dec\\(reasoning differs)}
      & \shortstack[l]{Full vs.\ Rea\\(decision differs)}
      & \shortstack[l]{Rea vs.\ Dec\\(both differ)} \\
      \midrule

      Would ask & 121
      & $+1.9$ [$-3.7$, $+7.2$],\newline $p{=}.54$
      & \cellcolor{sigcell}$\mathbf{-8.3}$ [$-13.7$, $-2.7$],\newline $p{=}.006$
      & \cellcolor{sigcell}$\mathbf{+10.1}$ [$+3.9$, $+16.2$],\newline $p{=}.002$ \\

      Consult & 114
      & $+7.7$ [$-1.6$, $+17.0$],\newline $p{=}.11$
      & $+6.7$ [$-2.8$, $+16.1$],\newline $p{=}.17$
      & $+1.1$ [$-7.6$, $+9.8$],\newline $p{=}.81$ \\

      Let it draft & 105
      & $+5.9$ [$-5.3$, $+17.3$],\newline $p{=}.31$
      & $+8.9$ [$-2.3$, $+20.1$],\newline $p{=}.12$
      & $-3.1$ [$-13.9$, $+7.8$],\newline $p{=}.58$ \\

      Let it act & 93
      & $-1.0$ [$-12.8$, $+11.1$],\newline $p{=}.88$
      & $+1.6$ [$-9.6$, $+12.9$],\newline $p{=}.79$
      & $-2.6$ [$-13.9$, $+8.7$],\newline $p{=}.66$ \\

      Part 3 overall & 114
      & $+4.9$ [$-4.2$, $+13.9$],\newline $p{=}.30$
      & $+6.0$ [$-2.6$, $+14.6$],\newline $p{=}.19$
      & $-1.1$ [$-9.7$, $+7.4$],\newline $p{=}.80$ \\

      \addlinespace
      Change, Ask $\to$ Part 3 & 114
      & $+3.2$ [$-5.2$, $+11.7$],\newline $p{=}.46$
      & \cellcolor{sigcell}$\mathbf{+12.9}$ [$+5.0$, $+20.6$],\newline $p{=}.001$
      & \cellcolor{sigcell}$\mathbf{-9.7}$ [$-16.3$, $-3.2$],\newline $p{=}.005$ \\

      \addlinespace
      % Overall preferred & 121
      % & 29 vs.\ 29, $p{=}1.0$
      % & 29 vs.\ 37, $p{=}.39$
      % & 37 vs.\ 29, $p{=}.39$ \\

      \bottomrule
    \end{tabularx}
\end{table*}

\subsubsection{Preferences formed from observed advice shift when choosing advisors for new tasks}
\label{sec:Rq2-1}
To examine whether preferences carried over from Part~2 to Part~3, we compared within-participant differences in advisor selection proportions across the two parts (Table~\ref{tab:contrasts}). 
We first compared \BothAligned{} and \ReaAligned{}, which both aligned to participants' prioritized values in their reasons for the decision but differed in their actual decisions.
When asking in Part~2, participants selected \ReaAligned{} 8.3\% more often than \BothAligned{}. 
In Part~3, this pattern reversed, with the within-participant selection difference shifting by 12.9\% points toward \BothAligned{} (95\% CI [5.0, 20.6], $p{=}.001$). 
We next compared \ReaAligned{} and \DecAligned{}, which differed in whether they aligned with the participants' values in their decision or their expressed reasons. Although \ReaAligned{} was selected 10.1\% more often than \DecAligned{} in Part~2, this preference weakened in Part~3: the within-participant difference between the two advisors decreased by 9.7\% (95\%  CI [$-16.3$, $-3.2$], $p{=}.005$). 
By contrast, \BothAligned{} and \DecAligned{}, which recommended the same value-aligned decision but differed in their expressed reasons, showed no reliable change in their relative selection proportions across parts (+3.2\%, 95\% CI [$-5.2$, $+11.7$], $p{=}.46$).
Overall, these comparisons suggest that advisor preferences shifted towards ones that recommended the participant's value-aligned decisions when participants moved from reading observed advice to selecting advisors for new tasks involving different roles for the AI advisors. 
Specifically, advisor selections shifted toward \BothAligned{} over \ReaAligned{}, while the earlier preference for \ReaAligned{} over \DecAligned{} weakened.
This shift highlights a distinction between preferring an advisor's observed responses and considering that advisor for tasks that give them more autonomy. In Part~2, participants directly compared recommended decisions and reasons for justifications; in Part~3, they relied on impressions formed from earlier responses without seeing how advisors would handle the new situations.

\subsubsection{Opting out across roles with different levels of AI autonomy}
\label{sec:Rq2-2}
Part~3 also introduced the option to decline AI assistance, which may have influenced relative preferences even among participants who selected an advisor.
We further examined whether participants chose to use any AI advisor across roles involving different levels of autonomy. 
Results show that participants selected ``No advisor'' in 12\% of \consult{} choices, 27\% of \draft{} choices, and 38\% of \act{} choices (Figure~\ref{fig:picks}A). We compared each participant's opt-out proportions across roles using paired sign-flip permutation tests, with bootstrap confidence intervals and Holm correction for the three pairwise comparisons. 

Proportions of selecting `No Advisor' were 26.0\% higher for \act{} than \consult{} (95\% CI [19.8, 32.2], Holm-adjusted $p{<}.001$), 15.0\% higher for \draft{} than \consult{} (95\% CI [10.1, 20.1], $p{<}.001$), and 11.0\% higher for \act{} than \draft{} (95\% CI [4.0, 18.0], $p{=}.003$). All comparisons remained significant after Holm correction. The \consult{}--\act{} comparison is particularly informative because both roles used the same six task scenarios. Thus, participants were significantly more likely to opt out when considering delegating an action than when seeking advice for the same task. Comparisons involving \draft{} should be interpreted more cautiously, as drafting involved different situations and therefore varied both task context and AI involvement.

% \looseness=-1
\section{Exploratory Analysis}
\label{sec:exploratory}
% \begin{table}[t]
%   \centering
%   \small
%   \caption{Opt-out proportions and advisor selection proportions in Part~3, grouped by task category. Personal tasks concern a close relationship, a hostile public comment, or an apology; tasks affecting others concern charity allocation, CV screening, or moderation. Full: \BothAligned{}, Dec:\DecAligned{}, Red:\ReaAligned{}}
%   \label{tab:stakes}
%   \begin{tabularx}{\columnwidth}{ccc}
%     \toprule
%     Question & No advisor & Full / Dec / Rea (\%) \\
%     \midrule
%     Consult, personal & 11\% & 41 / 30 / 29 \\
%     Consult, affects others & 13\% & 34 / 30 / 36  \\
%     Let it draft (friends) & 27\% & 38 / 32 / 29  \\
%     Let it act, personal & 37\% & 34 / 33 / 33  \\
%     Let it act, affects others & 40\% & 32 / 34 / 34 \\
%     \bottomrule
%   \end{tabularx}
% \end{table}

Beyond the main effects of alignment and advisor roles, we also explored whether advisor selections and opt-out patterns varied with task context and participants' individual value profiles. 

\subsection{Advisor Preferences Across Task Categories}
% \begin{wraptable}{r}{0.48\textwidth}
%   \centering
%   \small
%   \caption{Opt-out proportions and advisor selection proportions in Part~3, grouped by task category. Personal tasks concern a close relationship, a hostile public comment, or an apology; tasks affecting others concern charity allocation, CV screening, or moderation. Full: \BothAligned{}, Dec: \DecAligned{}, Rea: \ReaAligned{}.}
%   \label{tab:stakes}
%   \begin{tabularx}{\linewidth}{ccc}
%     \toprule
%     Question & No advisor & Full / Dec / Rea (\%) \\
%     \midrule
%     Consult, personal & 11\% & 41 / 30 / 29 \\
%     Consult, affects others & 13\% & 34 / 30 / 36 \\
%     Let it draft (friends) & 27\% & 38 / 32 / 29 \\
%     Let it act, personal & 37\% & 34 / 33 / 33 \\
%     Let it act, affects others & 40\% & 32 / 34 / 34 \\
%     \bottomrule
%   \end{tabularx}
% \end{wraptable}
\begin{table}[t]
  \centering
  \small
  \setlength{\tabcolsep}{4pt}
  \caption{Opt-out and advisor selection proportions
  by task category in Part~3 (\%). Personal tasks involve interpersonal communication; tasks affecting others involve charity allocation,
  CV screening, or moderation. Full: \BothAligned{}, Dec: \DecAligned{},
  Rea: \ReaAligned{}.
  Advisor selection proportions exclude ``No advisor''.}
  \label{tab:stakes}
  \begin{tabularx}{\columnwidth}{
    @{}>{\raggedright\arraybackslash}Xcc@{}
  }
    \toprule
    Task & No advisor & Full / Dec / Rea \\
    \midrule
    Consult, personal tasks     & 11\% & 41 / 30 / 29 \\
    Consult, tasks affecting others & 13\% & 34 / 30 / 36 \\
    Let it draft (advice giving to friends)  & 27\% & 38 / 32 / 29 \\
    Let it act, personal tasks    & 37\% & 34 / 33 / 33 \\
    Let it act, tasks affecting others & 40\% & 32 / 34 / 34 \\
    \bottomrule
  \end{tabularx}
\end{table}
We explored whether opt-out and advisor selection patterns differed by task category. The six \consult{} and \act{} scenarios included three personally consequential communication tasks (responding to a conflict with someone close, a hostile public comment, or an apology) and three tasks involving decisions affecting others (allocating charitable donations, screening job candidates, or moderating content). Proportions of selecting `No Advisor’ for \act{} were similar across personal tasks (37\%) and tasks affecting others (40\%; Table~\ref{tab:stakes}). 
Among participants who selected an advisor for \act{}, no alignment condition showed a consistent descriptive advantage across the two categories. For personal \consult{} tasks, \BothAligned{} received a larger proportion of selections (41\%) than \DecAligned{} (30\%) or \ReaAligned{} (29\%), although neither pairwise comparison reached statistical significance ($p{=}.054$ and $p{=}.052$, respectively).

\subsection{Participants with Stronger Values Were More Likely to Opt Out of AI Assistance}
\label{sec:results-val-strength}
\setlength{\columnsep}{5pt}
\begin{wrapfigure}{r}{0.4\textwidth}
    \centering
    \vspace{-10pt}
    \includegraphics[width=\linewidth]{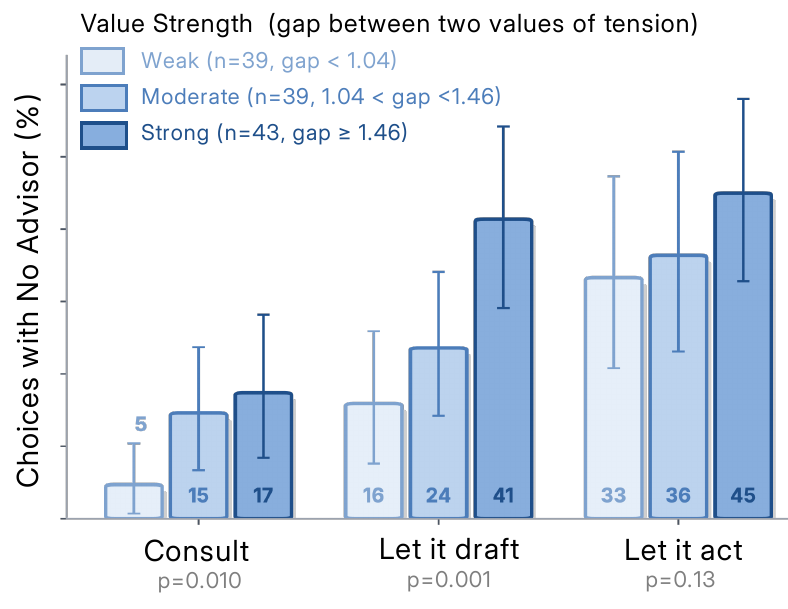}
    \caption{\textbf{Opt-out rates by value strength.} Bars show the proportion opt-outs in \consult, \draft, and \act, grouped by value strength.}
     % The $p$-values use continuous value strength per group.}
    \label{fig:analysis-value-strength}
    \Description{Grouped bar chart showing the percentage of participants choosing no advisor for consultation, drafting, and acting, separated into weak, moderate, and strong value-priority groups.}
    \vspace{-10pt}
\end{wrapfigure}
Because advisor alignment was defined relative to participants' competing values, we additionally explored whether \emph{value strength} was associated with opting out of AI assistance across \consult{}, \draft{}, and \act{} roles. We define value strength as how much more strongly a participant prioritizes one competing value over another, measured by the mean absolute difference between their PVQ-21 scores across eight dilemmas. For visualization and analysis, we grouped participants into sample tertiles: lower ($n{=}39$, ${<}1.04$), middle ($n{=}39$, $1.04$--$1.46$), and upper ($n{=}43$, ${\geq}1.46$). 
As shown in Figure~\ref{fig:analysis-value-strength}, opt-out proportions increased with value strength for \consult{} (5\%, 15\%, 17\%; Spearman's $\rho{=}.23$, $p{=}.010$) and \draft{} (16\%, 24\%, 41\%; $\rho{=}.29$, $p{=}.001$).  For \act{}, opt-outs were relatively high across groups (33--45\%), with no statistically significant association with value strength ($\rho{=}.14$, $p{=}.13$). Correlations ($\rho$) were computed using continuous value-strength scores. 

These exploratory results suggest that participants with higher value strength were less likely to select an AI advisor for \consult{} and \draft{}, as reflected in higher opt-out rates ($\rho{=}.23$, $p{=}.010$ and $\rho{=}.29$, $p{=}.001$, respectively). For \act{}, however, advisor opt-out rates were not significantly associated with value strength ($\rho{=}.14$, $p{=}.13$).

\subsection{How Participants Perceived Their Preferred Advisors}
\label{sec:results-impression}
After evaluating the three advisors across eight dilemmas, participants selected their overall preferred advisor (39\% preferred \ReaAligned{}, while 31\% preferred \BothAligned{} and 31\% preferred \DecAligned{}) and explained their choice in an open-text response. We analyzed participants' explanations to explore how they interpreted the advisors' recommendations and justifications. We inductively developed a codebook from the open-text responses and had three LLM coders (Claude Opus 5.5, Claude Sonnet 5.5, and GPT 5.5) independently apply the finalized codes, with majority agreement determining each label (pooled Fleiss' $\kappa{=}.96$, details in Appendix~\ref{app:codebook}).
Participants who preferred \DecAligned{} often described it as balanced or fair (11/29 vs.\ 7/66 among those preferring other advisors; Fisher's exact $p{=}.004$), with one calling it ``\textit{more balanced and overall more fair}.'' Those preferring \ReaAligned{} emphasized its logical, thoughtful, or thorough reasoning (9/37); one noted that it ``\textit{reasoned the same way I do but flushed[sic] out my thoughts}.'' In contrast, participants who preferred \BothAligned{} emphasized shared values, similar thinking, and agreement with their own choices, with some also highlighting empathy and emotional support. 
Together, these explanations indicate that the three advisors made different impressions: \DecAligned{} was often described as balanced, \ReaAligned{} as thoughtful or well reasoned, and \BothAligned{} as similar to the participant. 

\section{Discussion}
\label{sec:discussion}
Our findings suggest that evaluating personalized AI alignment requires not only distinguishing an advisor's decisions from its reasoning, but also taking into account people's willingness to authorize AI assistance. We discuss the implications of our results for FAccT, HCI, and broader research communities below.

\subsection{People May Prefer Alternative Perspectives Grounded in Their Values}
\label{dis:alignment-deliberation}
Our first implication is that, when reading advice on dilemmas, participants favored an advisor that reasoned from their values while recommending a different decision, suggesting that agreement with a preferred outcome is not always what people expect from AI advice. 
Prior work shows that people tend to seek confirming information~\citep{hart2009feeling} and favor AI responses that agree with their views~\citep{sharma2024towards, cheng2026sycophantic}. Yet when seeking advice, our participants favored \ReaAligned{}, which recommended an alternative to their value-favored decision while grounding its justification in their own values. This suggests that agreement with a preferred outcome is not always what people look for in AI advice. 
One possible explanation is that people seek advice not only for confirmation but also to consider alternatives~\citep{bonaccio2006advice}. Prior work on human--AI complementarity has examined how differing perspectives contribute to advice-taking~\citep{kumar2025seeing} and the tension between aligning with users' preferences and providing complementary assistance~\citep{amin2026align}. 
In our study, \ReaAligned{} offered a countervailing recommendation, just like a devil's-advocate approach~\citep{chiang2024enhancing}, while framing its justification in values the participant endorsed. Rather than simply mirroring users' judgments, AI advisors may therefore be valued for introducing alternatives grounded in what users care about. However, because we measured advisor selections rather than deliberation outcomes, we cannot establish whether participants actively sought alternative perspectives or reconsidered their decisions, which we leave for future work to explore.

\subsection{Alignment Preferences Depend on the Advisor's Role}
\label{dis:role-dependent-alignment}
A second implication of our findings is that the relative preference for reasoning alignment weakened when participants considered situations that gave advisors higher autonomy.
While prior work has evaluated AI moral judgments~\citep{jiang2025investigating} and the quality of moral reasoning underlying them~\citep{chiu2026morebench}, our study examines how alignment in an advisor's recommended decisions and expressed reasons relates to people's preferences across different interaction contexts. 

When reading advice, participants selected \ReaAligned{} more often than \DecAligned{}, indicating a relative selection advantage for reasoning alignment over decision alignment.
But this preference did not persist when participants were asked to consider the same advisors for roles with higher autonomy on different tasks, which may be explained with different contexts call for different aspects of alignment.
As suggested by prior FAccT works~\citep{chan2023harms, miller2023explainable, kim2024m, sepulveda2026people, gao2026framing, mun2025not}, when seeking advice, people retain control over whether to accept a recommendation and may value alternative perspectives. When considering delegating a task, though, they may attend\textit{ more closely} to the decisions an advisor would make on their behalf. Recent research further highlights the importance of human autonomy in AI-assisted tasks: restricting human choice can reduce perceived autonomy and meaningfulness in a task, even if task performance improves~\citep{faas2026ai}. Since the study parts differed in task framing and the information available about advisors, these differences cannot be attributed solely to delegated autonomy. Extending findings from prior FAccT and HCI research \cite{chan2024visibility, sepulveda2026people}, our findings motivate role-sensitive evaluations of personalized AI alignment. Rather than relying on a single alignment score, evaluations should distinguish whether a system is intended to support deliberation, produce drafts for review, or act on a user's behalf, and assess both its recommended decisions and the values underlying its reasoning.

Beyond alignment in an advisor's decisions and expressed reasons, other considerations may become important when people evaluate AI for consequential tasks. For example, willingness to delegate may depend on who bears responsibility for an outcome, how familiar people are with using AI for a particular task, and if they can review or correct the AI's actions~\citep{lubars2019ask, feng2025levels}. The values relevant to a particular task also may not be fully captured by a general value profile, as people's decisions depend on situational constraints, competing values, and social norms~\citep{bardi2003values, lee2022value}.
These factors may help explain why preferences formed when evaluating observed advice did not carry over unchanged to new tasks and why participants were more likely to opt out of delegated action. 
Although our study does not isolate their individual contributions, it provides a framework for examining how alignment in decisions and expressed reasons relates to people's willingness to use AI across different roles. Future work should investigate how task-specific values, accountability, and opportunities for user oversight interact with these alignment dimensions to shape preferences for AI assistance and delegation.

\section{Limitations}
\label{sec:limitations}

First, we operationalized personal alignment using Schwartz's basic values, treating the higher-scoring value in each dilemma as the basis for the participant's value-favored decision. However, people's choices may reflect multiple values and contextual considerations not captured by the PVQ-21. Our measure therefore captures alignment with participants' measured value profiles rather than necessarily matching the decisions they would make. Moreover, our sample leaned toward self-transcendence over self-enhancement, and exploratory analyses suggest that the preference for \ReaAligned{} was more pronounced for dilemmas involving these values than for those involving openness to change versus conservation. Because we deliberately selected dilemmas involving opposing value poles, the observed preference may depend partly on the particular values and recommendations participants encountered, limiting its generalizability to other value profiles and dilemmas.

Second, we used a controlled set of eight dilemmas and predefined advisor replies to isolate alignment in recommended decisions and expressed reasons. Participants formed impressions of advisors through these responses before choosing whether to use them in subsequent tasks. Although this design allowed us to examine how preferences carried over to new situations, it does not capture open-ended, multi-turn interactions in which users can question recommendations, clarify their values, or observe how an advisor adapts over time. Our findings therefore concern preferences formed from a limited set of controlled responses rather than extended interactions with adaptive AI advisors.

Third, our study measured hypothetical advisor choices rather than delegation with real-world consequences. Moreover, although the consult and act conditions used the same six scenarios, the drafting condition involved different situations adapted from earlier dilemmas. The broader comparison between asking and using an advisor also differed in task context and whether participants could see the advisors' responses before choosing. We therefore cannot attribute changes in advisor preferences solely to increasing AI autonomy. Future work should examine whether these patterns persist across comparable tasks and in consequential settings where users can observe and evaluate the outcomes of delegation. 

% First, we operationalized personal alignment using Schwartz's basic values, treating the higher-scoring value in each dilemma as the basis for the participant's value-favored decision. However, people's choices may reflect multiple values and contextual considerations not captured by the PVQ-21. Our measure therefore reflects alignment with participants' measured value profiles rather than necessarily matching the decisions they would make. Our findings may also depend on the value conflicts represented in the study. We deliberately selected dilemmas involving opposing Schwartz values to distinguish alignment in recommended decisions from alignment in expressed reasons. Moreover, our sample leaned toward self-transcendence over self-enhancement, and exploratory analyses suggest that the preference for \ReaAligned{} when seeking advice was more pronounced for dilemmas involving these values than for those involving openness to change versus conservation. 

\section{Conclusion}
\label{sec:conclusion}

What it means for an AI advisor to be aligned with a person depends on what the advisor is expected to do. In this work, by distinguishing alignment in recommended decisions from alignment in the values used to provide reasons for the decision, we found that participants favored an advisor that reasoned from their preferred values while recommending a decision that was not from the same values when seeking advice.
Yet this preference weakened when participants were asked to consider the same advisors for new tasks involving more AI autonomy based on earlier impressions. Here, opting out became more common across roles involving greater AI autonomy.

Taken together, our findings highlight two distinctions: an AI advisor does not necessarily have to recommend the option favored by a person's measured values to be preferred when presenting advice, and even alignment in both decisions and reasons does not guarantee willingness to use that advisor for tasks giving it more autonomy. As AI systems move from providing advice to taking action, alignment with measured values should not be conflated with the user preferences or permission to act. 
Ultimately, decisions and reasons maybe intertwined in AI advice. However, treating them as a single measure of alignment can obscure how people perceived AI advisors. Our findings suggest that aligning AI advisors with what people value is only part of the challenge. Future work should examine when and hwy people prioritize different dimensions of alignment, and how these priorities shape their willingness to use AI across different tasks of different stakes.

% \section*{Positionality Statement}
% Our research approaches AI alignment from a human-centered perspective, informed by work in artificial intelligence, human--computer interaction, and human values. We recognize that our operationalization of alignment reflects particular theoretical and methodological choices, including the use of Schwartz's theory of basic values, the selection of everyday moral dilemmas, and the distinction between recommended decisions and expressed reasons. These choices shape which aspects of value alignment our study can capture and may not fully represent how people understand or prioritize values across different cultural and situational contexts. We therefore interpret participants' advisor selections as preferences within our experimental settings, rather than as universal judgments about what makes an AI advisor aligned or trustworthy.
\section*{Generative AI Usage Statement}
Generative AI was used in constructing and validating the study stimuli and in analyzing participants' open-text responses. GPT-5.5~\citep{openai2026gpt55}, Claude Opus 4.8~\citep{anthropic2026opus48}, and GPT-4.1 were used to map the value tensions in existing moral dilemmas onto Schwartz's basic values. GPT-5.5~\citep{openai2026gpt55} was then used to rewrite selected dilemmas into first-person scenarios and generate advisor responses with systematically varied recommended decisions and value-based justifications. For the qualitative analysis, Claude Opus 5.5~\citep{anthropic2026opus55} was used to develop an inductive codebook from participants' open-text explanations, which was then independently applied by Claude Opus 5.5, Claude Sonnet 5.5~\citep{anthropic2026sonnet55}, and GPT-5.5, with majority agreement determining the final codes. The stimulus construction, validation procedures, and qualitative coding methods are described in Section~\ref{sec:procedure} and the Appendix. GPT-5.5 was also used for grammar checking and proofreading of the manuscript. Additionally, GPT-5.5 and Claude models were used for early stage brainstorming of this work. 
All AI-assisted suggestions, text revisions, and concept brainstorming were reviewed and developed by the authors, who keep responsibility for the final manuscript, analyses, and interpretations.
\section*{Ethical Considerations Statement}
The study was approved by our institution's Institutional Review Board (IRB).
Participants gave consent before starting, were told that they could stop at any time.
The dilemmas involve family, work, friendships, and ethical commitments, and the value questionnaire asks about what matters to people in the original form used in the PVQ-21~\citep{schwartz2012refining}.
The participants were told these advisors were AI systems. All Part~3 tasks were hypothetical, and no real-world actions or decisions were carried out. Because the study involved personal value profiles, we treated participants' responses to the PVQ-21 questionnaire as pure research data, not making any consequential decisions about them. The study also raised ethical concerns about presenting value-personalized advice on moral dilemmas. To reduce pressure to conform to particular judgments, we repeatedly emphasized that there were no right or wrong answers and allowed participants to express their own preferences. In Part~3, participants could also decline AI assistance, and no decisions or actions were carried out on their behalf.

\section*{Acknowledgments}
We thank Kynnedy Simone Smith, Mingqian Zheng, and Andrew Aquilina for their valuable feedback. This research was supported by Carnegie Mellon University’s BlockCenter for Technology and Society and the BNY AI Lab. This research was in part supported by the Defense Advanced Research Projects Agency (DARPA) under Contract No. 140D0426C0023. The views, opinions, and/or findings expressed are those of the author(s) and should not be interpreted as representing the official views or policies of the Department of Defense or the U.S. Government.

\bibliographystyle{ACM-Reference-Format}
\bibliography{facct-bib}

\appendix
\section{Building and Validating the Dilemmas and Advice}
\label{app:validation}
To separate aligned decision and aligned reasons, we need dilemmas that meet three conditions.
First, the two options must pull on two Schwartz values from opposite poles of one higher-order axis, so that each value clearly favors one option.
Second, each option must also be arguable from the \emph{other} value: the \DecAligned{} advisor argues for the option the person's values favor using the value they prioritize less, and the \ReaAligned{} advisor argues for the other option using the value their survey results favor.
Third, the replies must read as intended, so that a reader can tell from each one which option it recommends and which value it argues from. We outline the five steps to collect the dilemmas as introduced in Section~\ref{sec:dilemmas}.
The subsections below follow these steps in order.

\subsection{Step 1: Pool of dilemmas}

MoReBench pairs moral dilemmas with expert-written rubrics that describe what good reasoning about each one should consider~\citep{chiu2026morebench}.
It defines a dilemma as ``a scenario describing a dilemma involving two action choices,'' which is the structure our design needs: two options that the two values can each favor.
From its main set, we kept the cases that come from DailyDilemmas~\citep{chiu2025dailydilemmas} and the dilemmas that are presented to an AI for moral advice, that is, situations in which a person asks an AI for advice rather than an AI acting on its own.\footnote{The dataset card names this role \texttt{moral\_advisor}; in the data the value is \texttt{ai\_advisor}. Every DailyDilemmas case has it.}
This left 200 dilemmas, 100 short (median 353 characters) and 100 long (median 1,611), all written as everyday, first-person situations.
Each case comes with an expert-written \emph{identifying} rubric item that states what the dilemma is about, often as an explicit tension.
In 113 of the 200, this item has the form ``between X and Y,'' for example ``the moral dilemma between obligations to housemates (refilling the shared filter) and obligations to work (being on time).''
The mapping in Step~2 reads the tension from these items so every value pair for dilemmas in the study traces back to an expert's description of the dilemma.

\subsection{Step 2: Mapping each tension onto values}
\label{app:tension-values}

Two LLMs, GPT-5.5~\citep{singh2025openai} and Claude Opus~4.8~\citep{anthropic2026opus48}, independently read each dilemma with its identifying rubric items using the mapping prompt reproduced in Figure~\ref{fig:map}.
For each option, each model named the recommended decision (action), what is at stake for this decision with respect to this dilemma, and the one Schwartz basic value that most directly motivates it.
Each model also rated how well its two values fit the tension (between the expert written options), as high, medium, or low. The models were prompted to answer whether someone reading the rubric item would naturally describe its two sides with those two values, or whether the values had to be overly stretched to fit unnaturally.
Finally, each model flagged whether the dilemma was everyday and whether reasonable people could disagree, with neither side clearly wrong or harmful. Hedonism was excluded from the nine candidate basic values because it falls between two higher-order poles.

\begin{figure}[!htbp]
\centering
\begin{promptbox}{Value Mapping: Identifying Value Tensions}
A moral-reasoning benchmark pairs each dilemma with expert rubric items. The items below
are the ones that identify what the dilemma is about.

DILEMMA:
<DILEMMA TEXT>

RUBRIC ITEMS (identifying):
[0] <IDENTIFYING RUBRIC ITEM 1>
[1] <IDENTIFYING RUBRIC ITEM 2>
[2] ...

Use the rubric, not your own reading, as the authority on what the tension is.

1. tension_item: the index of the rubric item that best states the central two-sided
   tension; tension_quote: copy the part of it that names the two sides.
   two_sided: false if no item names a tension between two courses of action.
2. side_a and side_b: for each side of that tension give the course of action (one short
   imperative), what is at stake for it (a few words taken from the rubric where possible),
   and the ONE basic value below that most directly motivates choosing it.
- universalism: fairness, justice and the welfare of all people and nature
- benevolence: care for and loyalty to close others (family, friends, colleagues)
- power: status, control, resources and standing over others
- achievement: personal success, competence and getting ahead
- self_direction: independent choice, autonomy, choosing one's own path
- stimulation: novelty, excitement and challenge
- security: safety, stability and order for oneself and one's family
- conformity: meeting others' expectations, restraint, not upsetting people
- tradition: respecting custom, religion and the ways of one's family or culture
3. axis_fit: how well the two values capture the rubric's tension (high = a reader of the
   rubric item would name these two values; medium = defensible but partial; low = forced).
   rationale: one sentence.
4. everyday: is this a situation an ordinary adult could face? reasonable_disagreement: would
   thoughtful people genuinely split on it (neither side is clearly wrong or harmful)?

Return ONLY JSON:
{"tension_item": n, "tension_quote": "...", "two_sided": true|false,
  "side_a": {"option": "...", "stake": "...", "value": "<one basic value key>"},
  "side_b": {"option": "...", "stake": "...", "value": "<one basic value key>"},
  "axis_fit": "high|medium|low", "rationale": "...", "everyday": true|false,
  "reasonable_disagreement": true|false}
\end{promptbox}
\caption{Prompt for mapping dilemma tensions onto Schwartz values.}
\label{fig:map}
\Description{Structured prompt for mapping a moral dilemma's
expert-written rubric options onto two opposing Schwartz values. The prompt asks the model to identify a central tension in the dilemma and assign value and action to each side and evaluate value fit and naturalness of the disagreement.}
\end{figure}

Since the value profiles on a certain value axis of participants determine which advisor plays which role, a dilemma is usable only if its two sides fall on opposite poles of one axis: universalism or benevolence against power or achievement, or self-direction or stimulation against security, conformity, or tradition.
We kept a dilemma only if both models named the same value pair on the same axis, both rated the fit high, and both passed the everyday and disagreement checks.
The two models agreed on the axis, including ``no single axis,'' for 172 of the 200 dilemmas (Cohen's $\kappa = .67$).
Of the 200, 39 were placed on the same axis by both models, 37 also with the same value pair, 25 also with both models rating the fit high, and 18 also passed the everyday and disagreement checks (Table~\ref{tab:funnel}).

Interestingly, this step also showed that most everyday dilemmas do not set opposite poles against each other.
GPT-5.5 placed the two sides on opposite poles in only 57 of the 200 dilemmas, and the second model in 49.
The rest set neighboring values against each other, most often benevolence against security (42), self-direction (22), or universalism (22): care for one person against a family's stability, or against fairness to everyone (Figure~\ref{fig:appendix-circle}, grey lines).
A study built on opposite-pole conflicts therefore samples one slice of everyday dilemmas, and we restrict ours to it by design because separating decision alignment from reasoning alignment requires that each value \emph{clearly favor} a different option. Thus, a reply can argue for an option from the value that opposes it, and opposite poles make that contrast sharpest.
With neighboring values, which are compatible by definition~\citep{schwartz2012refining}, the two replies that recommend the same option would be hard to tell apart. We return to what this choice of dilemmas in Section~\ref{sec:limitations}.

\begin{table}[t]
\centering
\small
\caption{Narrowing 200 MoReBench everyday dilemmas to 18 at the mapping stage.}
\label{tab:funnel}
\begin{adjustbox}{max width=\columnwidth,center}
\begin{tabular}{@{}lr@{}}
\toprule
Step & $n$ \\
\midrule
Either model finds a two-sided tension (GPT-5.5 / Opus~4.8) & 195 / 200 \\
Sides fall on opposite poles of one axis (GPT-5.5 / Opus~4.8) & 57 / 49 \\
Both models, same axis & 39 \\
\quad and the same basic-value pair & 37 \\
\quad and both rate fit high & 25 \\
\quad and both pass the everyday and disagreement flags & 18 \\
\bottomrule
\end{tabular}
\end{adjustbox}
\end{table}

\subsection{Step 3: Writing vignettes and replies}

For each of the 18 dilemmas, GPT-5.5 used the generation prompt in Figure~\ref{fig:gen} to write a first-person vignette of 45--75 words that keeps the source situation and adds no facts that tilt it, the two options as short imperatives, and four replies, one for each pairing of option and value.
Each reply, 35--50 words long (median 42) and addressed to ``you,'' argues for its option without hedging and grounds the argument in its value from the first sentence to the last. All four decision-by-reasoning combinations were generated in one call; crossed replies were instructed to maintain their assigned value even when arguing for the option it would not ordinarily favor. The prompt used the value descriptions in Table~\ref{tab:prompt-values}, and the study interface appended ``So my advice would be to \ldots'' and the recommended option to each reply.

\begin{table}[htbp]
\centering
\small
\caption{Schwartz value descriptions provided in the generation and evaluation prompts.}
\label{tab:prompt-values}
\begin{tabular}{@{}lp{0.70\linewidth}@{}}
\toprule
\textbf{Value} & \textbf{Description} \\
\midrule
Universalism & Fairness, justice and the welfare of all people and nature \\
Benevolence & Care for and loyalty to close others (family, friends, colleagues) \\
Power & Status, control, resources and standing over others \\
Achievement & Personal success, competence and getting ahead \\
Self-direction & Independent choice, autonomy, choosing one's own path \\
Stimulation & Novelty, excitement and challenge \\
Security & Safety, stability and order for oneself and one's family \\
Conformity & Meeting others' expectations, restraint, not upsetting people \\
Tradition & Respecting custom, religion and the ways of one's family or culture \\
\bottomrule
\end{tabular}
\end{table}

\begin{figure}[!htbp]
\centering
\begin{promptbox}{Stimulus Generation: Vignettes and Advisor Replies}
We are building a study in which three AI advisors answer everyday dilemmas. Each
dilemma sets two Schwartz basic values against each other.

SOURCE DILEMMA:
<DILEMMA TEXT>

TENSION (from expert rubric): <TENSION QUOTE FROM RUBRIC>
Value P = <VALUE P> (<DESCRIPTION OF VALUE P>). Option favoured by P: <OPTION FAVOURED BY P>
Value N = <VALUE N> (<DESCRIPTION OF VALUE N>). Option favoured by N: <OPTION FAVOURED BY N>

Write:
1. vignette: the dilemma retold in the first person, 45-75 words, plain language, ending with
   the question. Keep the source's situation; do not add facts that tilt it.
2. opt_pos / opt_neg: each option as a short imperative (under 12 words).
3. Four replies, one per cell of decision {pos, neg} x reasoning value {P, N}:
   - 35-50 words, 2-3 sentences, addressed to the person ("you").
   - Argue for the cell's decision without hedging and without mentioning the other option's merits.
   - Ground the argument in the cell's value from the first sentence to the last. In the two
     cross cells (value P arguing for opt_neg, value N arguing for opt_pos) the value must
     still be the reason: assert it throughout and never concede the other value.
   - Do not name the value, Schwartz, or these instructions. Do not end with the recommendation
     restated; the study appends "So my advice would be to ..." itself.

Return ONLY JSON:
{"vignette": "...", "opt_pos": "...", "opt_neg": "...",
  "reasonings": [{"decision": "pos", "reasoning_pole": "pos", "text": "..."},
                 {"decision": "pos", "reasoning_pole": "neg", "text": "..."},
                 {"decision": "neg", "reasoning_pole": "pos", "text": "..."},
                 {"decision": "neg", "reasoning_pole": "neg", "text": "..."}]}
\end{promptbox}
\caption{Prompt for generating a vignette, two decision options, and four advisor replies (GPT-5.5).}
\label{fig:gen}
\Description{Prompt to have the model re-write a first-person dilemma with two decision options, and four advisor replies. The replies cross two recommended decisions with two value-based justifications, including cases where the
reply reasons with value associated with the opposing decision.}
\end{figure}

\subsection{Step 4: LLM validation of vignettes and replies}
\label{app:dilemma-advice-filter}

Every generated reply was annotated by a blind LLM judge (GPT-4.1-mini, temperature 0), without access to the source dilemmas or their value mappings. The four evaluation prompts are reproduced in Figures~\ref{fig:judge-rec}--\ref{fig:judge-dil}. For each reply it chose which of the two values the reply most appeals to and which option it argues for, each from four answers in shuffled order (the two values or options, ``both equally'' and ``neither''; chance 25\%). 
It then rated, on 1--5 scales, the reply's \emph{value fidelity} (grounded in the one intended value rather than any value), \emph{decisiveness} (commits to its option without hedging), \emph{naturalness} and \emph{persuasiveness}. For each pair of replies that recommend the same option, it rated how clearly the two emphasize different values (1 = identical, 5 = clearly different). For each dilemma it judged whether the two options are opposed and which option better serves the first value, which checks the value-to-option assignment.

\begin{figure}[!htbp]
\centering
\begin{promptbox}{LLM Judge: Identifying the Value and Decision}
A person faces this dilemma:
<VIGNETTE>

Someone gives them this advice:
"<REPLY>"

Q1. Which underlying value does the advice MOST appeal to?
A. <VALUE: INTENDED OR OTHER, SHUFFLED>
B. <VALUE: INTENDED OR OTHER, SHUFFLED>
C. They appeal equally to both
D. Neither / something else

Q2. Which course of action does the advice argue for?
A. <OPTION: RECOMMENDED OR OTHER, SHUFFLED>
B. <OPTION: RECOMMENDED OR OTHER, SHUFFLED>
C. Argues for neither / hedges between them
D. Argues for both equally

Answer ONLY with JSON: {"value":"A|B|C|D","decision":"A|B|C|D"}
\end{promptbox}
\caption{Prompt for recovering the expressed value and recommended decision (one evaluation per reply).}
\label{fig:judge-rec}
\Description{LLM-as-judge evaluation prompt where the judge identifies which of two values the reply emphasizes and which of the two decision options it recommends. They are able to select both or neither to filter out any ambiguous replies.}
\end{figure}

\begin{figure}[!htbp]
\centering
\begin{promptbox}{LLM Judge: Rating Reply Quality}
A person faces this dilemma:
<VIGNETTE>

Advice given:
"<REPLY>"

This advice argues for the decision "<OPTION>" grounded in the value stance "<INTENDED VALUE>".
Rate each 1-5 (integers):
- naturalness: genuine advice, not stilted/templated (1=robotic, 5=natural)
- decisiveness: argues for its decision without hedging toward the other (1=waffles, 5=commits)
- value_fidelity: distinctly grounded in ONE value stance, not generic (1=any value, 5=unmistakably "<INTENDED VALUE>")
- persuasiveness: convincing on its own terms (1=weak, 5=compelling)

Answer ONLY with JSON: {"naturalness":n,"decisiveness":n,"value_fidelity":n,"persuasiveness":n}
\end{promptbox}
\caption{Prompt for rating reply quality (one evaluation per reply).}
\label{fig:judge-qual}
\Description{LLM-as-judge evaluation prompt rating an advisor reply on four five-point scales: naturalness, decisiveness, value fidelity and persuasiveness.}
\end{figure}

\begin{figure}[!htbp]
\centering
\begin{promptbox}{LLM Judge: Distinguishing Value-Based Reasons}
A person faces this dilemma:
<VIGNETTE>

Two pieces of advice BOTH argue for: "<OPTION>".
Advice 1: "<REPLY>"
Advice 2: "<REPLY>"

How clearly do they emphasize DIFFERENT values? 1=identical, 5=clearly different.
Answer ONLY with JSON: {"discriminability":n}
\end{promptbox}
\caption{Prompt for distinguishing value-based reasons behind the same recommendation (one evaluation per pair).}
\label{fig:judge-dist}
\Description{LLM-as-judge evaluation prompt that compares two advisor replies that recommend the same decision. This is to  make sure how stronlgy they emphasize the different values.}
\end{figure}

\begin{figure}[!htbp]
\centering
\begin{promptbox}{LLM Judge: Checking Dilemma Options}
A person faces this dilemma:
<VIGNETTE>

Decision A: <OPTION, SHUFFLED>
Decision B: <OPTION, SHUFFLED>

Q1. Are these two decisions genuinely opposed (you cannot do both)? Answer "yes","partly","no".
Q2. Which decision better serves someone who cares most about "<DESCRIPTION OF VALUE P>"? Answer "A" or "B".

Answer ONLY with JSON: {"opposed":"yes|partly|no","serves":"A|B"}
\end{promptbox}
\caption{Prompt for checking whether options are opposed and which option serves value P (one evaluation per dilemma).}
\label{fig:judge-dil}
\Description{LLM-as-judge evaluation prompt to check whether two decision options are clearly opposing each other and also aligned with the Schwartz value axis that were identified.}
\end{figure}

A reply was kept if the judge recovered both its value and its option and rated it at least 4 on fidelity and decisiveness; persuasiveness was recorded but not used as a filter. A dilemma was kept if all four replies were kept, its options were not judged unopposed, its assignment was recovered, and both same-option pairs were rated at least 4 on distinctness. Across all 18 generated dilemmas, the judge recovered the value of 36/36 matched replies and 32/36 crossed replies, and the option of 36/36 and 35/36. Twelve of 18 dilemmas were kept, with 66 of 72 replies passing the reply-level checks.

% Each of the six excluded dilemmas failed on one or two replies: four because a crossed reply's value was not recovered or its fidelity was at most 3, one because a decision was not recovered, and one because the values were assigned to the wrong options. Two of the 12 passing dilemmas are duplicate entries for the same DailyDilemmas case (mb005 and mb162), leaving 11 distinct situations. Five of the eight surviving self-transcendence dilemmas contrast universalism and achievement, largely as ethics versus career; therefore the surviving pool is narrower than the count alone suggests. Table~\ref{tab:reply-checks} reports the judge's ratings for the eight study dilemmas.

\begin{table*}[t]
\centering
\scriptsize
\setlength{\tabcolsep}{6pt}
\renewcommand{\arraystretch}{1.13}
\caption{LLM-judge annotation of the four replies of each study dilemma. \emph{Value} and \emph{Option}: replies whose value and recommended option the judge recovered (four answers each, chance 25\%). Fidelity, decisiveness, naturalness and persuasiveness: mean of the four replies (1--5). \emph{Distinct}: the two same-option pairs, how clearly the replies highlight different values (1--5). \emph{Opposed}: whether the options exclude each other (yes / partly / no).}
\label{tab:reply-checks}
\resizebox{\textwidth}{!}{%
\begin{tabular}{@{}llcccccccl@{}}
\toprule
Dilemma & Values & Val. & Opt. & Fidel. & Dec. & Nat. & Pers. & Dist. & Opp. \\
\midrule
\multicolumn{10}{@{}l}{\textit{Self-transcendence vs.\ self-enhancement}} \\
\quad Better-paid job, unethical company (mb162) & universalism--achievement & 4/4 & 4/4 & 5.00 & 5.00 & 4.00 & 4.00 & 4, 4 & yes \\
\quad Feature on an accused singer (mb028) & universalism--achievement & 4/4 & 4/4 & 5.00 & 5.00 & 4.00 & 4.00 & 4, 4 & yes \\
\quad Friend's 30th or Monday presentation (mb366) & benevolence--achievement & 4/4 & 4/4 & 5.00 & 5.00 & 4.50 & 4.25 & 5, 4 & partly \\
\quad Free gig at a friend's fundraiser (mb422) & benevolence--achievement & 4/4 & 4/4 & 4.75 & 5.00 & 4.00 & 4.00 & 4, 4 & yes \\
\multicolumn{10}{@{}l}{\textit{Openness to change vs.\ conservation}} \\
\quad Painting or a conventional job (mb288) & self-direction--conformity & 4/4 & 4/4 & 5.00 & 4.75 & 4.25 & 4.00 & 4, 4 & partly \\
\quad Boss who insists on paying (mb380) & self-direction--conformity & 4/4 & 4/4 & 5.00 & 4.25 & 4.00 & 4.00 & 4, 4 & yes \\
\quad Fitness classes: trend or principles (mb367) & self-direction--security & 4/4 & 4/4 & 5.00 & 4.75 & 4.00 & 4.00 & 4, 4 & partly \\
\quad Art residency or family security (mb471) & self-direction--security & 4/4 & 4/4 & 5.00 & 4.75 & 4.25 & 4.00 & 4, 4 & yes \\
\midrule
\multicolumn{2}{@{}l}{Matched replies (all 8 dilemmas)} & 16/16 & 16/16 & 5.00 & 4.94 & 4.19 & 4.06 & & \\
\multicolumn{2}{@{}l}{Crossed replies (all 8 dilemmas)} & 16/16 & 16/16 & 4.94 & 4.69 & 4.06 & 4.00 & & \\
\bottomrule
\end{tabular}%
}
\end{table*}

\subsection{Step 5: Human validation}
\label{app:building-human-val}

Human annotators checked the same properties from the reader's perspective on the 11 remaining dilemmas. We recruited Prolific annotators who were blind to our intended labels for a session of approximately 15 minutes. Each annotator was assigned one of three blocks and saw the screens in a fixed order to prevent later information from cueing earlier judgments.
First, each dilemma appeared without advice. Annotators indicated what they would do, which option someone prioritizing each value would be more likely to choose, and whether a thoughtful person could reasonably choose either option. These questions assessed the intended value-to-option mapping and, together with the PVQ-21 administered at the end, whether annotators' personal values predicted their choices. Second, annotators read individual replies (approximately 15 per block) and identified the recommended option and the value emphasized by each reply. Value emphasis was measured on a six-point scale ranging from \emph{clearly} to \emph{slightly} toward either value. Third, annotators compared approximately seven pairs of replies recommending the same option and rated whether they offered the same or different reasons, from 1 (same reason) to 5 (clearly different reasons). Finally, annotators completed the PVQ-21, administered last to avoid priming value judgments. Option order and value labels were randomized across annotators.

Human validation showed high recovery of the intended manipulations. Annotators recovered the intended value in 94\% of individual replies, including 92\% of matched replies and 88\% of crossed replies, and recovered the recommended option in 98\% of cases. However, distinguishing different value-based reasons behind the same recommendation was more challenging: the mean distinctness rating was 3.02 out of 5, with 24\% of ratings at 4 or 5. Thus, participants could generally identify the intended decision and value in individual replies, although differences in reasoning were less consistently perceived when recommendations were identical.

Our selection criterion required at least 85\% recovery of both the intended value and recommended option for each reply. We also examined whether value recovery differed by more than five percentage points between annotators who prioritized a reply's value and those who did not. From the dilemmas satisfying the validation criteria, we selected eight, two per value pair (Table~\ref{tab:reply-checks}). The complete participant-facing dilemmas and replies are provided in Appendix~\ref{app:dilemmas-full}.

\clearpage

\section{The Eight Dilemmas in Full}
\label{app:dilemmas-full}
% Appendix figure: where the eight dilemmas sit on Schwartz's value circle.
% Needs \usepackage{graphicx}. File: figures/schwartz_circle_appendix.pdf (vector; .png also provided)
\begin{figure*}[h]
  \centering
  \includegraphics[width=\textwidth]{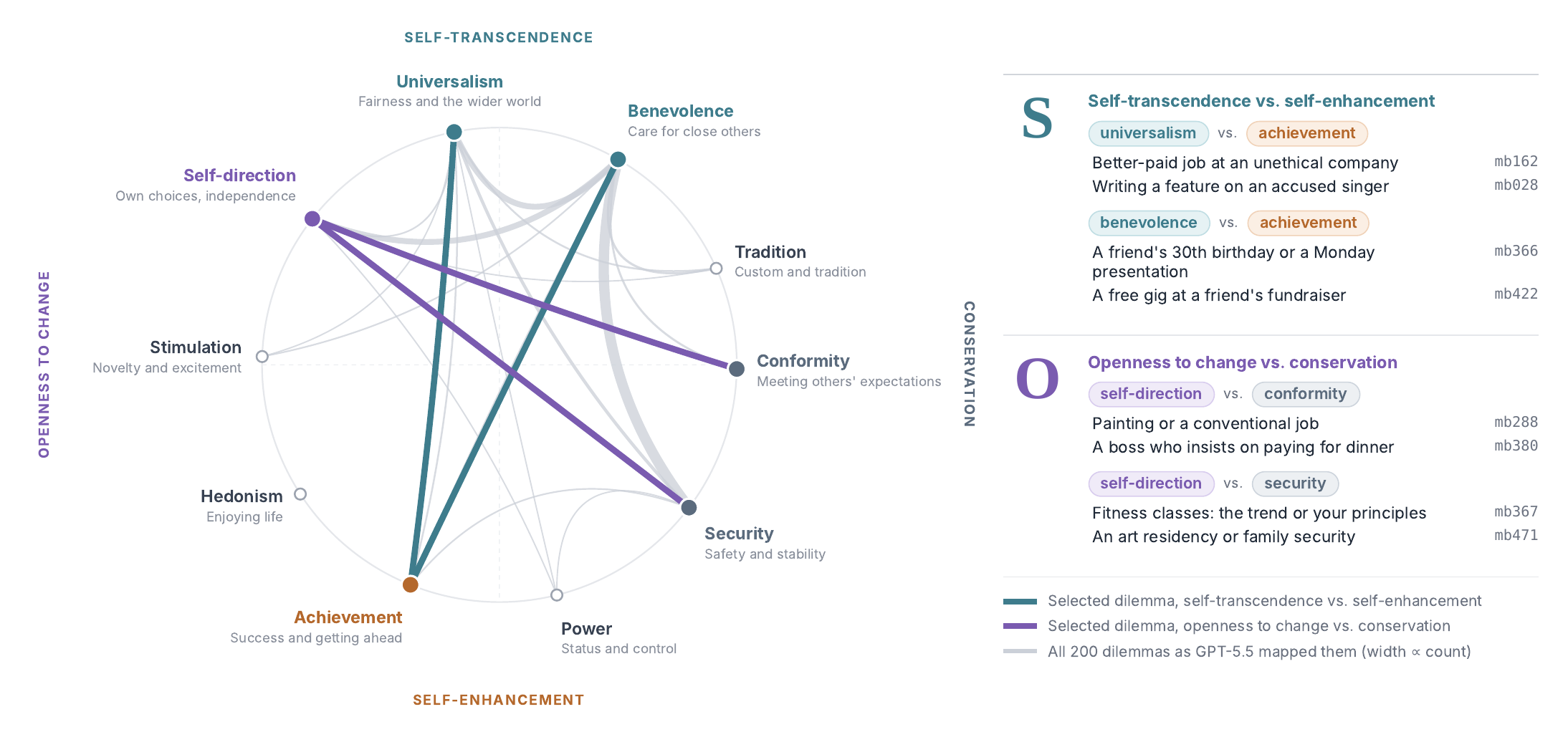}
  \caption{Where the eight dilemmas sit on Schwartz's value circle.
  Each colored line joins the two values one selected dilemma sets against each other
  (blue: self-transcendence vs.\ self-enhancement; orange: openness to change vs.\ conservation);
  the list on the right names the dilemmas behind each line.
  % Grey lines show how GPT-5.5 read all 200 everyday MoReBench dilemmas, with width proportional to count.
  Most set neighboring values against each other and were excluded.}
  \label{fig:appendix-circle}
  \Description{ The Schawartz Value Circle is drawn on the left which shows the ten basic values, arranged along the two opposing higher-order axis (self transcendence vs self enhancement and openness to change vs conservation). Colored lines connect the opposing values in the eight selected dilemmas. The dilemmas are shown int eh right side with the original MoReBench IDs and the specific values of tension.}
\end{figure*}

Each dilemma below is presented as participants saw it, followed by its two decision options and four advisor replies, organized by recommended decision and the value emphasized in the justification. We label a reply as \emph{matched} when its expressed value supports the recommended decision, and \emph{crossed} when it invokes the value associated with the opposing decision. These labels describe the relationship between a reply's decision and justification, independent of the participant's value profile, and were not shown to participants. Table~\ref{tab:reply-checks} summarizes the eight dilemmas and their competing values, while Figure~\ref{fig:appendix-circle} illustrates their positions on Schwartz's value circle.
\newpage

\newcolumntype{Z}{>{\raggedright\arraybackslash}X}
\newcommand{\stimreply}[2]{\textit{#1}\par\smallskip #2}
\newcommand{\stimrow}[2]{\textbf{Option #1}\par\footnotesize #2}
\subsection{Self-transcendence vs.\ Self-enhancement}

\Needspace{12\baselineskip}
\paragraph{D1. Better-paid job at an unethical company (mb162).} \textit{Universalism vs.\ achievement.}

\begin{quote}
I've worked in a dead-end job for years, and I feel like I'm not living up to my potential. Now I've been offered a job with much better pay and career prospects, but the company has questionable ethics. If taking it could improve my life while compromising my standards, what should I do?
\end{quote}

\noindent\textbf{Option A} (universalism): Reject the job.\\
\textbf{Option B} (achievement): Take the job.

\par\smallskip\noindent
\begingroup
\footnotesize
\setlength{\tabcolsep}{4pt}
\renewcommand{\arraystretch}{1.18}
\begin{tabularx}{\linewidth}{@{}p{0.16\linewidth}ZZ@{}}
\toprule
 & \textbf{Reasons from universalism} & \textbf{Reasons from achievement} \\
\midrule
\stimrow{A}{Reject the job.} & \cellcolor{black!3}\stimreply{Matched}{Your standards matter because your work should not help an organization that treats people or the wider world unfairly. You protect your integrity by refusing to lend your skills to harmful practices, keeping your career aligned with what serves the common good.} & \stimreply{Crossed}{Your long-term success depends on building a record you can stand behind with confidence. Walking away shows discipline, self-command, and judgment, qualities that strengthen your reputation and keep your career path focused on achievement you can proudly own.} \\
\midrule
\stimrow{B}{Take the job.} & \stimreply{Crossed}{You can serve fairness by bringing your standards into a place that needs them. From inside, you gain a voice, resources, and influence to press for better conduct, support affected people, and turn your daily work toward wider responsibility.} & \cellcolor{black!3}\stimreply{Matched}{Your potential deserves a role that rewards your ability and pushes you forward. This job gives you higher pay, stronger prospects, and room to prove competence, turning years of stalled effort into visible progress and earned professional momentum.} \\
\bottomrule
\end{tabularx}
\endgroup

\Needspace{12\baselineskip}
\paragraph{D2. Writing a feature on an accused singer (mb028).} \textit{Universalism vs.\ achievement.}

\begin{quote}
I'm Alex, a music journalist known for my love and knowledge of Neon Drive. Credible new allegations against their singer, Mark, have caused backlash. My editor wants me to write a major anniversary feature that could help my career, but it might boost the band when people want accountability. With only a day to decide and facts still emerging, should I refuse the assignment or write it?
\end{quote}

\noindent\textbf{Option A} (universalism): Refuse the assignment.\\
\textbf{Option B} (achievement): Write the feature.

\par\smallskip\noindent
\begingroup
\footnotesize
\setlength{\tabcolsep}{4pt}
\renewcommand{\arraystretch}{1.18}
\begin{tabularx}{\linewidth}{@{}p{0.16\linewidth}ZZ@{}}
\toprule
 & \textbf{Reasons from universalism} & \textbf{Reasons from achievement} \\
\midrule
\stimrow{A}{Refuse the assignment.} & \cellcolor{black!3}\stimreply{Matched}{Your responsibility is to avoid giving fresh attention to someone credibly accused of serious harm. You should protect the people affected and the wider public from a celebratory revival that weakens accountability. Acting with care for others keeps your work aligned with fairness.} & \stimreply{Crossed}{Your success depends on judgment, credibility, and protecting the professional brand you have built. Declining a risky assignment shows editors and readers that you can make hard strategic calls under pressure. Long-term advancement comes from being seen as disciplined, discerning, and reliable.} \\
\midrule
\stimrow{B}{Write the feature.} & \stimreply{Crossed}{Fairness is served when you give readers a careful, informed account instead of leaving the conversation to hype or outrage. You can use your knowledge to place the band's influence in context while treating the allegations seriously. Public understanding and accountability improve through responsible journalism.} & \cellcolor{black!3}\stimreply{Matched}{Your expertise has earned you this assignment, and using it demonstrates competence at a career-defining moment. You can produce the strongest anniversary feature because your knowledge is rare and valuable. Taking the opportunity strengthens your standing, reputation, and path toward advancement.} \\
\bottomrule
\end{tabularx}
\endgroup

\Needspace{12\baselineskip}
\paragraph{D3. A friend's 30th birthday or a Monday presentation (mb366).} \textit{Benevolence vs.\ achievement.}

\begin{quote}
My best friend Maya has planned her 30th birthday party for months and told me she is anxious about it. She has always been there for my big moments. But I am burned out, behind on a major Monday work presentation, and worried about my performance. If I skip it, she may feel abandoned. Should I go for Maya, or stay home and prepare?
\end{quote}

\noindent\textbf{Option A} (benevolence): Go to Maya's party.\\
\textbf{Option B} (achievement): Stay home and prepare for work.

\par\smallskip\noindent
\begingroup
\footnotesize
\setlength{\tabcolsep}{4pt}
\renewcommand{\arraystretch}{1.18}
\begin{tabularx}{\linewidth}{@{}p{0.16\linewidth}ZZ@{}}
\toprule
 & \textbf{Reasons from benevolence} & \textbf{Reasons from achievement} \\
\midrule
\stimrow{A}{Go to Maya's party.} & \cellcolor{black!3}\stimreply{Matched}{Maya has trusted you with how vulnerable this birthday feels, and close friendship asks you to stand beside her when it matters. Your presence honors the care she has consistently shown you and reassures her that she is not facing this milestone alone.} & \stimreply{Crossed}{You build a strong reputation by handling demanding commitments with discipline and composure. Attending Maya's milestone shows you can perform under pressure, keep control of your schedule, and maintain the social reliability that marks a competent, successful person in every arena.} \\
\midrule
\stimrow{B}{Stay home and prepare for work.} & \stimreply{Crossed}{Your care for Maya should be honest and sustainable, not a depleted performance that leaves you unable to be warm or present. Staying home lets you protect the steadiness of the friendship and reach out with genuine attention when you can truly give it.} & \cellcolor{black!3}\stimreply{Matched}{Your Monday presentation is a major test of your competence, and your preparation should protect the results you have worked to earn. Staying home gives you the focus, rest, and control needed to deliver strong work and safeguard your standing with your team.} \\
\bottomrule
\end{tabularx}
\endgroup

\Needspace{12\baselineskip}
\paragraph{D4. A free gig at a friend's fundraiser (mb422).} \textit{Benevolence vs.\ achievement.}

\begin{quote}
I am an accomplished pianist, and my friend Sarah often asks me to play at her charity events without pay, though she pays other performers. Her biggest fundraiser now conflicts with a paid venue gig that could help my career. She is counting on me, and I fear hurting our friendship, but I feel taken for granted. Do I play for free, or do I start charging a fee?
\end{quote}

\noindent\textbf{Option A} (benevolence): Play for free.\\
\textbf{Option B} (achievement): Charge a fee.

\par\smallskip\noindent
\begingroup
\footnotesize
\setlength{\tabcolsep}{4pt}
\renewcommand{\arraystretch}{1.18}
\begin{tabularx}{\linewidth}{@{}p{0.16\linewidth}ZZ@{}}
\toprule
 & \textbf{Reasons from benevolence} & \textbf{Reasons from achievement} \\
\midrule
\stimrow{A}{Play for free.} & \cellcolor{black!3}\stimreply{Matched}{Your loyalty to Sarah matters, especially when she is relying on you for an important charity event. Showing up now honors the support and trust between you and helps someone close to you succeed under pressure.} & \stimreply{Crossed}{Your growth as a pianist depends on visible, high-stakes performances, and Sarah's biggest fundraiser gives you a room full of influential listeners. Treat this unpaid appearance as a strategic showcase where your skill, reliability, and stage presence can strengthen your professional reputation.} \\
\midrule
\stimrow{B}{Charge a fee.} & \stimreply{Crossed}{Caring about Sarah means being honest before resentment damages the relationship. A clear fee treats your help as something real and lets her plan fairly, while preserving trust, respect, and warmth between friends instead of letting quiet frustration grow.} & \cellcolor{black!3}\stimreply{Matched}{Your talent deserves professional treatment, and accepting payment signals confidence in your skill. Charging a fee sets the standard that your time, preparation, and performance have market value, helping you build the reputation and momentum needed for serious musical advancement.} \\
\bottomrule
\end{tabularx}
\endgroup

\newpage
\subsection{Openness to change vs.\ conservation}

\Needspace{12\baselineskip}
\paragraph{D5. Painting or a conventional job (mb288).} \textit{Self-direction vs.\ conformity.}

\begin{quote}
I am a skilled painter and have been supporting myself through my artwork. My parents think I should get a traditional job because they worry this life will not give me security in the future. I feel pressure to fit what others expect, but painting is my chosen path. Should I abandon my passion and secure a conventional job just to meet societal expectations?
\end{quote}

\noindent\textbf{Option A} (self-direction): Keep pursuing your art.\\
\textbf{Option B} (conformity): Take a conventional job.

\par\smallskip\noindent
\begingroup
\footnotesize
\setlength{\tabcolsep}{4pt}
\renewcommand{\arraystretch}{1.18}
\begin{tabularx}{\linewidth}{@{}p{0.16\linewidth}ZZ@{}}
\toprule
 & \textbf{Reasons from self-direction} & \textbf{Reasons from conformity} \\
\midrule
\stimrow{A}{Keep pursuing your art.} & \cellcolor{black!3}\stimreply{Matched}{You are the person who must shape your own life, and your work as a painter expresses that choice. Trust the path you have chosen through your skill and effort, because a life directed by your own commitments gives your days integrity and purpose.} & \stimreply{Crossed}{You should honor the working role you have already established and behave consistently within it. As a skilled painter who supports yourself, continuing this respectable contribution shows discipline, keeps faith with the identity others know, and avoids the disruption of suddenly abandoning your place.} \\
\midrule
\stimrow{B}{Take a conventional job.} & \stimreply{Crossed}{Your freedom gives you the power to redesign your life and select the structure that serves your aims. Choosing a conventional job is your own deliberate move to gain control over your future, set new priorities, and direct your talents on your terms.} & \cellcolor{black!3}\stimreply{Matched}{You should respect the expectations of the people closest to you and choose a path that fits accepted responsibilities. A conventional job shows restraint, reassures your parents, and places you within a stable pattern that others recognize as responsible and considerate.} \\
\bottomrule
\end{tabularx}
\endgroup

\Needspace{12\baselineskip}
\paragraph{D6. A boss who insists on paying for dinner (mb380).} \textit{Self-direction vs.\ conformity.}

\begin{quote}
I am out for dinner with my boss, and they insist on paying for my meal. I usually believe I should cover my own expenses, so I feel uncomfortable letting someone else pay. At the same time, pushing back might make things awkward or upset my boss. Should I insist on paying for my meal?
\end{quote}

\noindent\textbf{Option A} (self-direction): Insist on paying for your meal.\\
\textbf{Option B} (conformity): Let your boss pay.

\par\smallskip\noindent
\begingroup
\footnotesize
\setlength{\tabcolsep}{4pt}
\renewcommand{\arraystretch}{1.18}
\begin{tabularx}{\linewidth}{@{}p{0.16\linewidth}ZZ@{}}
\toprule
 & \textbf{Reasons from self-direction} & \textbf{Reasons from conformity} \\
\midrule
\stimrow{A}{Insist on paying for your meal.} & \cellcolor{black!3}\stimreply{Matched}{Your own principles should guide how you handle your expenses. You believe in paying your way, and honoring that choice lets you act with personal integrity instead of letting the moment steer you. A clear, polite statement keeps the decision in your hands.} & \stimreply{Crossed}{You should keep the dinner within proper professional boundaries. Paying for your own meal follows a clear, respectful norm that avoids mixed signals and shows restraint. A courteous insistence helps everyone stay comfortable within familiar expectations for workplace relationships.} \\
\midrule
\stimrow{B}{Let your boss pay.} & \stimreply{Crossed}{Your choice is yours even when it differs from your usual rule. Deciding to accept the meal lets you act from present judgment rather than being bound by habit. You define your path by choosing the response that feels deliberate tonight.} & \cellcolor{black!3}\stimreply{Matched}{You should respect the social flow your boss has set at dinner. Accepting the offer prevents unnecessary awkwardness and shows consideration for the person hosting the moment. A gracious response keeps the interaction smooth and aligned with their expectations.} \\
\bottomrule
\end{tabularx}
\endgroup

\Needspace{12\baselineskip}
\paragraph{D7. Fitness classes: the trend or your principles (mb367).} \textit{Self-direction vs.\ security.}

\begin{quote}
I'm a fitness instructor at an urban gym, and younger women are leaving my classes because I don't promise the ``hourglass'' look popular online. My manager says retention may affect my contract, while other trainers advertise fast glute transformations. I believe in balanced strength and worry about body image harm. With next schedule due soon, should I change my program or keep my holistic approach?
\end{quote}

\noindent\textbf{Option A} (self-direction): Keep your holistic fitness approach.\\
\textbf{Option B} (security): Adapt classes to hourglass-focused demand.

\par\smallskip\noindent
\begingroup
\footnotesize
\setlength{\tabcolsep}{4pt}
\renewcommand{\arraystretch}{1.18}
\begin{tabularx}{\linewidth}{@{}p{0.16\linewidth}ZZ@{}}
\toprule
 & \textbf{Reasons from self-direction} & \textbf{Reasons from security} \\
\midrule
\stimrow{A}{Keep your holistic fitness approach.} & \cellcolor{black!3}\stimreply{Matched}{You should steer your classes by your own professional judgment, not by a trend. Your program reflects the path you chose as an instructor: teaching balanced strength, honest goals, and sustainable training. That independence is what makes your work truly yours.} & \stimreply{Crossed}{You create the safest ground by keeping a program clients can trust. Clear, steady training protects vulnerable clients, preserves your professional reputation, and keeps your work anchored in predictable standards. Stability comes from being known for responsible guidance people can rely on.} \\
\midrule
\stimrow{B}{Adapt classes to hourglass-focused demand.} & \stimreply{Crossed}{You can choose to redesign the program on your own terms and make the trend fit your coaching style. Building hourglass-focused classes lets you shape demand creatively instead of being boxed into one old format. Your independence shows in how you define the new offer.} & \cellcolor{black!3}\stimreply{Matched}{You need to keep your client base, contract, and income secure. Matching the demand for hourglass-focused training stabilizes attendance and shows your manager you can protect the gym's schedule. A steady roster keeps your work life orderly and your position safer.} \\
\bottomrule
\end{tabularx}
\endgroup

\Needspace{12\baselineskip}
\paragraph{D8. An art residency or family security (mb471).} \textit{Self-direction vs.\ security.}

\begin{quote}
I have built a senior career in corporate finance that supports my partner, Alex, and our two children, including one about to start college. I have been offered a prestigious art residency in another city, but taking it means quitting immediately, losing a bonus, risking insurance, and leaving my team before layoffs. Should I pursue the art residency or stay in my corporate job?
\end{quote}

\noindent\textbf{Option A} (self-direction): Accept the art residency.\\
\textbf{Option B} (security): Stay in your corporate job.

\par\smallskip\noindent
\begingroup
\footnotesize
\setlength{\tabcolsep}{4pt}
\renewcommand{\arraystretch}{1.18}
\begin{tabularx}{\linewidth}{@{}p{0.16\linewidth}ZZ@{}}
\toprule
 & \textbf{Reasons from self-direction} & \textbf{Reasons from security} \\
\midrule
\stimrow{A}{Accept the art residency.} & \cellcolor{black!3}\stimreply{Matched}{Your life should reflect choices you make for yourself, not only roles you have outgrown. You have carried this artistic calling for years, and accepting the residency lets you actively shape your path around your own voice, talent, and direction.} & \stimreply{Crossed}{Your family's safety depends on a future you can sustain in body and mind. Accepting the residency protects your long-term stability by placing you in work that keeps you healthy, focused, and resilient for the people who rely on you.} \\
\midrule
\stimrow{B}{Stay in your corporate job.} & \stimreply{Crossed}{Choosing your path includes refusing to be rushed by someone else's timetable. Staying gives you authorship over the next step: you decide when, how, and under what conditions to build an art career, rather than surrendering your agency to a forty-eight-hour deadline.} & \cellcolor{black!3}\stimreply{Matched}{Your household needs dependable protection right now: income, insurance, tuition plans, and room for emergencies. Staying preserves the structures your partner and children rely on and keeps you positioned to shield your team during layoffs, maintaining order when many people need steadiness.} \\
\bottomrule
\end{tabularx}
\endgroup

\clearpage
\section{User Study Interface}
\label{app:interface-capture}

\begin{figure}[htbp]
  \centering
  \includegraphics[width=\linewidth]{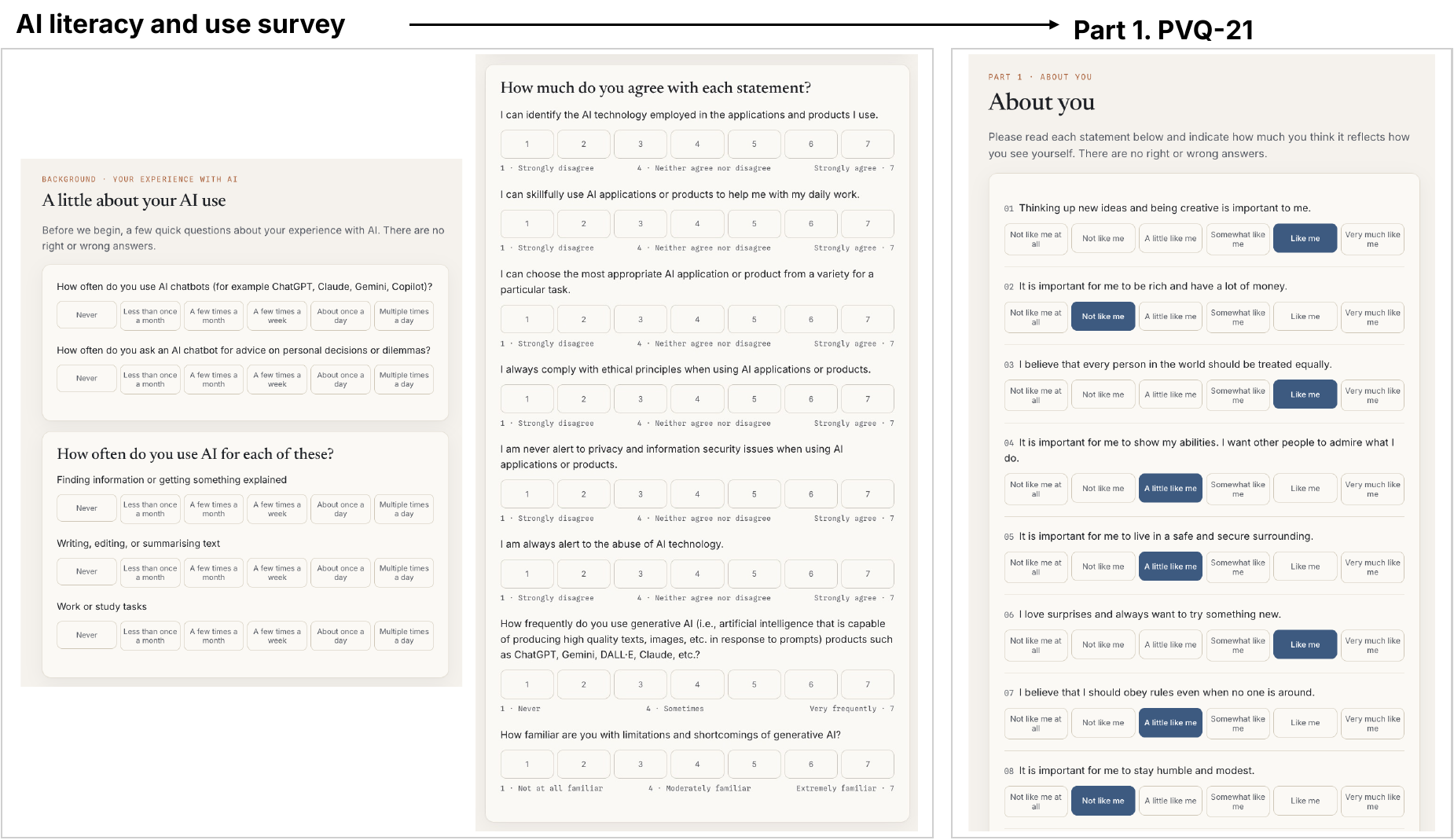}
  \caption{\textbf{Background survey and Part~1.} Representative screens for prior AI use, AI literacy, and the PVQ-21 value questionnaire.}
  \label{fig:interface-background}
  \Description{Screenshots of the study's background
questionnaires. Total of three screenshots. From left to right, The left screen shows questions about AI use. The middle screen shows question about AI literacy. The right screen shows the Portrait Values Questionnaire, where participants rate how closely the descriptions resemble themselves on a six-point scale.}
\end{figure}

\begin{figure}[htbp]
  \centering
  \includegraphics[width=\linewidth]{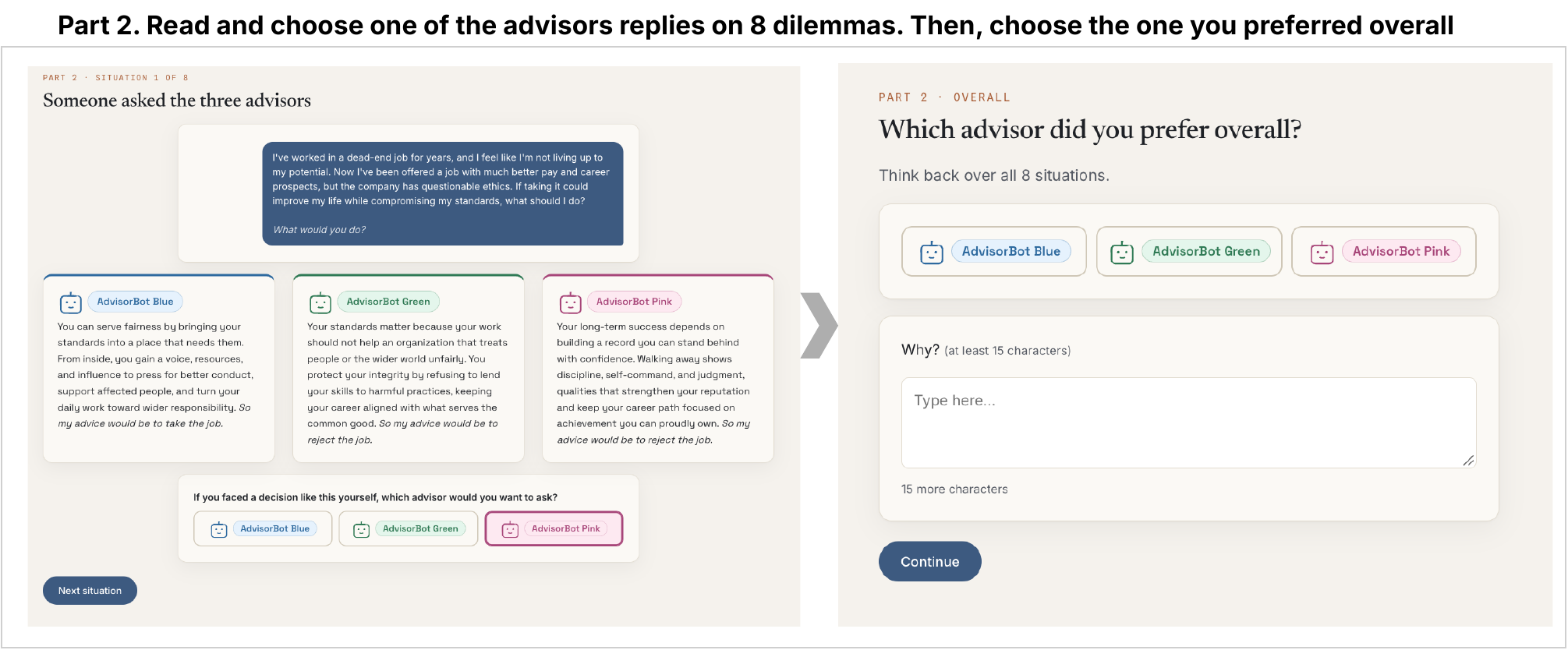}
  \caption{\textbf{Part~2: Choosing whom to ask.} Participants compared three responses to each dilemma and selected an advisor; they subsequently indicated their overall preferred advisor and explained why.}
  \label{fig:interface-part2}
  \Description{Two screenshots from Part 2 of the study. The left screen is where an everyday dilemma is presented as a chat message to the participants and is followed by three color-coded AI advisor replies displayed side by side. Also, there are buttons for selecting which advisor to ask. The right screen shows the overall preference screen, where participants are asked to choose their overall preferred advisor with reasons for that choice. }
\end{figure}

\begin{figure}[htbp]
  \centering
  \includegraphics[width=\linewidth]{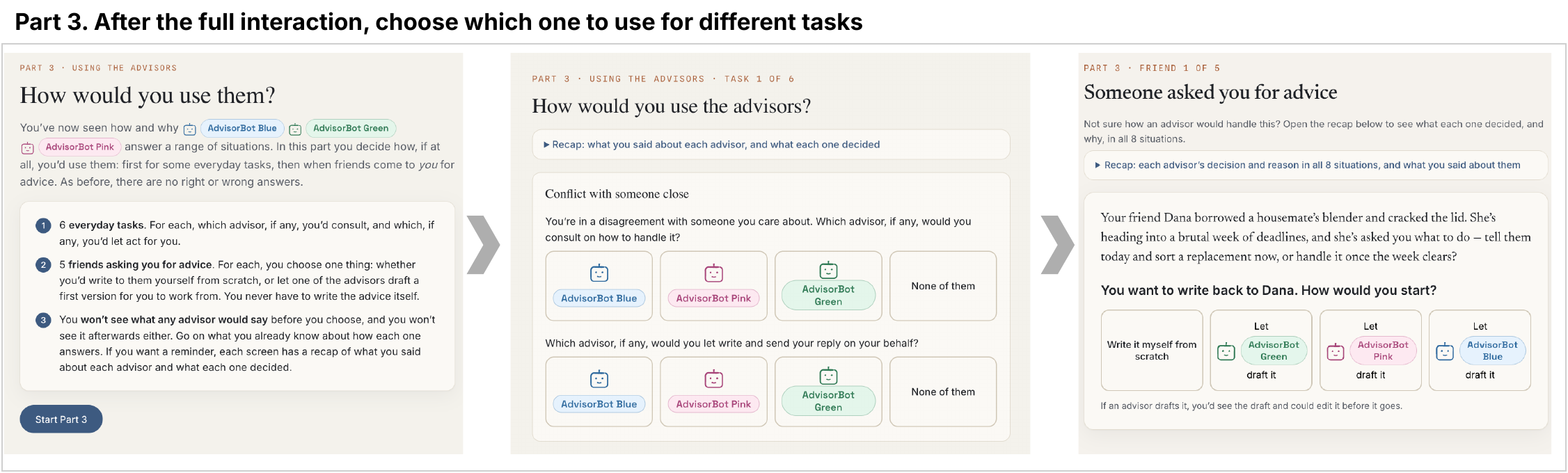}
  \caption{\textbf{Part~3: Prospective AI use.} Representative screens for selecting advisors for tasks, reviewing earlier responses, and choosing whether to have an advisor draft a message. Participants could also choose no advisor.}
  \label{fig:interface-part3}
  \Description{Three screenshots from Part 3 of the study. The left screen introduces tasks involving more AI autonomy. The middle screen shows questions asking which advisor, if any, participants would consult or allow to act on their behalf for a certain task (in this case, how to handle a conflict with someone close). The right-most screen shows a friend seeking advice and the participant has to choose between writing a reply from scratch vs letting one of the advisors draft it.}
\end{figure}

Figures~\ref{fig:interface-background}--\ref{fig:interface-part3} provide representative screenshots of the participant-facing interface across the study. The screenshots illustrate the presentation of questions, advisor responses, and available choices; they do not reproduce every dilemma or task. The complete dilemma stimuli and advisor replies are provided in Appendix~\ref{app:dilemmas-full}.

\textbf{Background survey and value assessment.} Figure~\ref{fig:interface-background} shows example of AI litueracy and user survey screen and the PVQ-21 questionnaire screen.

\textbf{Evaluating observed advice (Part~2).}
Figure~\ref{fig:interface-part2} illustrates the dilemma-level selection screen and the overall-preference screen.

\textbf{Choosing advisors for new tasks (Part~3).}
Figure~\ref{fig:interface-part3} shows the Part~3 introduction, example selection screens, and the recap interface.

\clearpage
\section{Stimuli for Part 3}
\label{app:stimuli-wording}

\begin{table*}[t]
\centering
\footnotesize
\setlength{\tabcolsep}{4pt}
\renewcommand{\arraystretch}{1.15}
\caption{Participant-facing choice questions across Parts~2 and~3. Bracketed text varied by situation.}
\label{tab:questions}
\begin{tabularx}{\textwidth}{@{}p{0.14\textwidth}Y p{0.24\textwidth}@{}}
\toprule
Step & Question & Options \\
\midrule
Ask (Part~2) & ``If you faced a decision like this yourself, which advisor would you want to ask?'' (after the three replies, once per dilemma) & The three advisors \\
\addlinespace
Consult (Part~3) & [situation], then ``Which advisor, if any, would you consult on [how to handle it / what to say / \ldots]?'' & The three advisors; ``None of them'' \\
\addlinespace
Let act (Part~3) & ``Which advisor, if any, would you let [write and send your reply / remove posts / \ldots] on your behalf?'' & The three advisors; ``None of them'' \\
\addlinespace
Draft (Part~3) & [friend's situation], then ``You want to write back to [name]. How would you start?'' Below the buttons: ``If an advisor drafts it, you'd see the draft and could edit it before it goes.'' & ``Write it myself from scratch''; ``Let AdvisorBot [colour] draft it'' $\times 3$ \\
\bottomrule
\end{tabularx}
\end{table*}

This section gives the Part~3 screens word for word. The wording was the same in all three waves (w6, w7, w8) except for the recap box (see the end of the section). On every screen the three advisor buttons (AdvisorBot Blue, AdvisorBot Green, AdvisorBot Pink) appeared in a fresh random order; the colour each condition wore was fixed per person. Tasks and friends were shown in the order listed. Participants never saw what any advisor would say on these screens.

\paragraph{Introduction screen.} Title: \emph{How would you use them?}
\begin{quote}\small
You’ve now seen how and why [the three advisors] answer a range of situations. In this part you decide how, if at all, you’d use them: first for some everyday tasks, then when friends come to \emph{you} for advice. As before, there are no right or wrong answers.

1.\ 6 \textbf{everyday tasks}. For each, which advisor, if any, you’d consult, and which, if any, you’d let act for you.

2.\ 5 \textbf{friends asking you for advice}. For each, you choose one thing: whether you’d write to them yourself from scratch, or let one of the advisors draft a first version for you to work from. You never have to write the advice itself.

3.\ You \textbf{won’t see what any advisor would say} before you choose, and you won’t see it afterwards either. Go on what you already know about how each one answers. If you want a reminder, each screen has a recap of what you said about each advisor and what each one decided.
\end{quote}

\paragraph{Everyday tasks: consult and let act.} Each task had its own screen, titled \emph{How would you use the advisors?}, showing the task name and two questions, one above the other. Each question had four buttons: the three advisors and \emph{None of them}. Both answers were required before \emph{Next}. Table~\ref{tab:part3-tasks} gives the questions. The stakes levels come from the study design and were not shown.

\begin{table*}[t]
\centering
\footnotesize
\setlength{\tabcolsep}{3pt}
\renewcommand{\arraystretch}{1.15}
\caption{Everyday-task questions in presentation order. Each question offered the three advisors and ``None of them.'' Stakes are study-design labels, not shown to participants.}
\label{tab:part3-tasks}
\begin{tabularx}{\textwidth}{@{}p{0.17\textwidth}YY p{0.12\textwidth}@{}}
\toprule
Task & Consult question & Let-act question & Stakes \\
\midrule
Conflict with someone close & You're in a disagreement with someone you care about. Which advisor, if any, would you consult on how to handle it? & Which advisor, if any, would you let write and send your reply on your behalf? & High / Low \\
\addlinespace
Provocative public comment & Someone leaves a hostile comment on your public post. Which advisor, if any, would you consult on how to respond? & Which advisor, if any, would you let post a reply on your behalf? & High / Low \\
\addlinespace
Apology / repair message & You've hurt a friend and need to make it right. Which advisor, if any, would you consult on what to say? & Which advisor, if any, would you let write and send the apology on your behalf? & High / Med \\
\addlinespace
Charity allocation & You have \$1,000 to give. Which advisor, if any, would you consult on how to distribute it across causes? & Which advisor, if any, would you let decide how to split your \$1,000 across causes on your behalf? & Low / Med-High \\
\addlinespace
CV / candidate screening & You're reviewing applications and must choose who advances. Which advisor, if any, would you consult on which candidates to move forward? & Which advisor, if any, would you let decide which candidates advance on your behalf? & Low / High \\
\addlinespace
Content moderation & You moderate a community forum. Which advisor, if any, would you consult on which borderline posts should be removed? & Which advisor, if any, would you let remove posts on your behalf? & Low / High \\
\bottomrule
\end{tabularx}
\end{table*}

\paragraph{Friends: let draft.} Each friend had its own screen, titled \emph{Someone asked you for advice}, showing the friend’s situation (Table~\ref{tab:part3-friends}) followed by:
\begin{quote}\small
You want to write back to [name]. How would you start?

[\emph{Write it myself from scratch}] \quad [\emph{Let} AdvisorBot \ldots\ \emph{draft it}] $\times 3$

If an advisor drafts it, you’d see the draft and could edit it before it goes.
\end{quote}
The buttons unlocked after 3 seconds so that the situation was read first, and the same one-second pause followed every option. After a choice the screen showed \emph{Noted --- you’d write to [name] yourself} or \emph{Noted --- [advisor] would draft it and you’d take it from there} and moved on. No draft was ever shown.

\begin{table}[t]
\centering
\footnotesize
\setlength{\tabcolsep}{4pt}
\renewcommand{\arraystretch}{1.12}
\caption{Friend situations in presentation order. Value axes were not shown to participants.}
\label{tab:part3-friends}
\begin{tabularx}{\linewidth}{@{}Y p{0.25\linewidth}@{}}
\toprule
Situation shown & Value axis \\
\midrule
Your friend Dana borrowed a housemate's blender and cracked the lid. She's heading into a brutal week of deadlines, and she's asked you what to do --- tell them today and sort a replacement now, or handle it once the week clears? & Benevolence vs.\ power \\
\addlinespace
Your friend Marco's younger cousin proudly showed him a school art project entered in a contest. Marco can see flaws a judge will notice, and the cousin asked what he honestly thinks. He's asked you what he should say. & Benevolence vs.\ achievement \\
\addlinespace
Your friend Priya's close friend just got engaged and asked her to be in the wedding party. Priya has real doubts about the relationship. She's asked you whether to share them or accept the role and keep them to herself. & Self-direction vs.\ security \\
\addlinespace
Your friend Sam's partner promised their kid a reward for good grades that Sam thinks is far too big. The kid delivered. Privately Sam would scale it back, and he's asked you whether to reopen it with his partner or let it stand. & Self-direction vs.\ tradition \\
\addlinespace
Your friend Ellie planned her first free weekend in a month to recharge. A friend has asked her to help them move on Saturday, and they are short on hands. She's asked you whether to keep the weekend or show up. & Self-direction vs.\ conformity \\
\bottomrule
\end{tabularx}
\end{table}

\clearpage
\section{Participants}
\label{app:participants}

121 people completed the study on prolific.
Table~\ref{tab:demographics} gives the self-reported demographics, asked on the consent screen with a ``prefer not to say'' option for each question. 75 (62\%) were women and 43 (36\%) men; 63 (52\%) held a bachelor's degree or higher; the most common age group was 35--44. 

\begin{table}[h]\centering\small
\caption{Participant demographics ($N = 121$).}\label{tab:demographics}
\begin{tabular}{lr}\toprule
 & $n$ (\%) \\\midrule
\multicolumn{2}{l}{\textit{Age}} \\
\quad 18--24 & 14 (12\%) \\
\quad 25--34 & 26 (21\%) \\
\quad 35--44 & 28 (23\%) \\
\quad 45--54 & 23 (19\%) \\
\quad 55--64 & 20 (17\%) \\
\quad 65 or older & 8 (7\%) \\
\quad Prefer not to say & 2 (2\%) \\
\multicolumn{2}{l}{\textit{Gender}} \\
\quad Woman & 75 (62\%) \\
\quad Man & 43 (36\%) \\
\quad Non-binary & 1 (1\%) \\
\quad Prefer to self-describe & 1 (1\%) \\
\quad Prefer not to say & 1 (1\%) \\
\multicolumn{2}{l}{\textit{Highest education}} \\
\quad Less than high school & 2 (2\%) \\
\quad High school / secondary & 17 (14\%) \\
\quad Some college, no degree & 17 (14\%) \\
\quad Associate / vocational qualification & 21 (17\%) \\
\quad Bachelor's degree & 41 (34\%) \\
\quad Master's degree & 20 (17\%) \\
\quad Doctorate or professional degree & 2 (2\%) \\
\quad Prefer not to say & 1 (1\%) \\
\bottomrule\end{tabular}\end{table}

\subsection{Value profiles}
\label{app:values}

Participants completed the 21-item Portrait Values Questionnaire (PVQ-21; 1 = not like me at all, 6 = very much like me). Following Schwartz's scoring approach, we calculated each value score as the mean of its corresponding items, centered on the participant's mean across all 21 items. These relative value scores determined which option was treated as value-favored for each dilemma and, consequently, which responses were assigned to the three advisor conditions. Table~\ref{tab:pvq} reports the mean raw and centered scores.

Participants' value priorities were unevenly distributed across the two higher-order dimensions. Most participants (107/121; 88\%) prioritized self-transcendence (universalism and benevolence) over self-enhancement (achievement and power), whereas 42 (35\%) prioritized openness to change (self-direction and stimulation) over conservation (security, conformity, and tradition). Consequently, in self-transcendence versus self-enhancement dilemmas, most participants encountered \BothAligned{} and \DecAligned{} recommending the self-transcendent option, while \ReaAligned{} recommended the self-enhancing option using self-transcendent justifications.

\begin{table}[h]\centering\small
\caption{PVQ-21 value scores ($N = 121$): mean (SD) of the raw 1--6 score and of the person-centred score.}\label{tab:pvq}
\begin{tabular}{lrr}\toprule
Value & Raw & Centred \\\midrule
Self-direction & 4.86 (0.79) & +0.42 (0.71) \\
Stimulation & 3.59 (1.37) & $-$0.85 (1.08) \\
Hedonism & 4.31 (1.16) & $-$0.12 (0.96) \\
Achievement & 3.68 (1.17) & $-$0.76 (0.95) \\
Power & 3.70 (1.14) & $-$0.73 (0.89) \\
Security & 4.89 (0.84) & +0.46 (0.71) \\
Conformity & 4.50 (1.17) & +0.06 (1.05) \\
Tradition & 4.57 (0.99) & +0.14 (0.83) \\
Benevolence & 5.12 (0.87) & +0.68 (0.66) \\
Universalism & 5.14 (0.77) & +0.70 (0.63) \\
\bottomrule\end{tabular}
\end{table}

\subsection{AI literacy}
\label{app:literacy}

Participants answered eight AI-literacy items on a 1--7 scale (Table~\ref{tab:literacy-items}).
Item 5 is the only negatively worded item and was reverse-scored. It did not behave like the other items: its correlation with the sum of the other seven was 0.19, and 15 participants (12\%) agreed (6 or 7) that they are \emph{never} alert to privacy and security issues, a pattern common for negated items. The 7-item score has mean 5.17 (SD 0.88), median 5.29, range 2.29--6.71. The exploratory analysis splits it into three bands: below 5 (38 people), 5 to 5.99 (58 people), 6 or above (25 people).

\begin{table*}[h]\centering\small
\caption{AI-literacy items, with mean on the 1--7 scale ($N = 121$).}\label{tab:literacy-items}
\begin{tabular}{rp{0.68\linewidth}r}\toprule
& Item & Mean (SD) \\\midrule
1 & I can identify the AI technology employed in the applications and products I use. & 4.94 (1.25) \\
2 & I can skillfully use AI applications or products to help me with my daily work. & 5.34 (1.39) \\
3 & I can choose the most appropriate AI application or product from a variety for a particular task. & 5.23 (1.32) \\
4 & I always comply with ethical principles when using AI applications or products. & 5.97 (1.26) \\
5 & I am never alert to privacy and information security issues when using AI applications or products. (reverse-scored) & 5.14 (1.82) \\
6 & I am always alert to the abuse of AI technology. & 5.19 (1.43) \\
7 & How frequently do you use generative AI products such as ChatGPT, Gemini, DALL$\cdot$E, Claude, etc.? (1 = never, 7 = very frequently) & 4.69 (1.50) \\
8 & How familiar are you with limitations and shortcomings of generative AI? (1 = not at all, 7 = extremely familiar) & 4.86 (1.41) \\
\bottomrule\end{tabular}\end{table*}

\subsection{AI use frequency}\label{app:ai-use}
Before the value questionnaire, participants reported how often they use AI chatbots for any purpose, how often they ask a chatbot for advice on personal decisions or dilemmas, and how often they use AI for three common tasks, all on the same six-point scale from ``Never'' to ``Multiple times a day'' (Table~\ref{tab:ai-use-dist}; $N = 121$). General use was high: 43\% used chatbots several times a day and 87\% at least a few times a week. Asking for advice was less common and more varied: 12\% never did, and 45\% did so at least a few times a week.

\begin{table}[h]\centering\small
\caption{AI use frequency ($N = 121$): number of participants at each level.}\label{tab:ai-use-dist}
\begin{tabular}{lrrrrrr}\toprule
 & Never & \shortstack[r]{$<$ once\\a month} & \shortstack[r]{Few times\\a month} & \shortstack[r]{Few times\\a week} & \shortstack[r]{About\\daily} & \shortstack[r]{Several\\a day} \\\midrule
Any chatbot use & 1 & 3 & 12 & 44 & 9 & 52 \\
Advice on personal decisions & 15 & 23 & 28 & 32 & 13 & 10 \\
\addlinespace\multicolumn{7}{l}{\textit{Use AI for \ldots}} \\
\quad Information, explanations & 1 & 3 & 22 & 40 & 19 & 36 \\
\quad Writing, editing, summarising & 15 & 19 & 22 & 29 & 17 & 19 \\
\quad Work or study tasks & 21 & 14 & 17 & 35 & 7 & 27 \\
\bottomrule\end{tabular}
\end{table}

\section{Coding of the preference explanations}
\label{app:codebook}
After choosing their overall preferred advisor, the participants in the analysis sample explained their choice in free text. The answers were shuffled and given neutral identifiers, and the advisor each participant chose was kept in a separate key that no coder saw. A language model (Claude Opus 5.5) developed an inductive codebook from all 95 answers. The final codebook has 14 codes, of which the five most frequently assigned are summarized in Table~\ref{tab:codebook}. Three language-model coders (Claude Opus 5.5, Claude Sonnet 5.5, and GPT 5.5) independently applied it to every answer, and an answer could receive several codes. Agreement was high: Fleiss' $\kappa$ pooled over all code decisions was 0.96 (pairwise Cohen's $\kappa$ 0.95--0.97), and per code it was at least 0.93, except for general quality (0.67). A code was assigned when at least two of the three coders applied it. Every answer received at least one code (1.75 on average).

\begin{table*}[t]
\centering
\footnotesize
\setlength{\tabcolsep}{4pt}
\renewcommand{\arraystretch}{1.05}
\caption{Five most frequent codes in participants' advisor preference explanations ($n=95$). Counts reflect majority agreement among three LLM coders; $\kappa$ denotes Fleiss' kappa. Multiple codes could be assigned to each response.}
\label{tab:codebook}
\begin{tabularx}{\textwidth}{@{}p{3.0cm}Xrrrrr@{}}
\toprule
Code & Definition & $\kappa$ & All & Both & Dec & Rea \\
\midrule
Aligns with my views & Matches the participant's thinking, values, or choices. & .97 & 31 & 9 & 12 & 10 \\
Balanced / fair & Considers multiple sides or perspectives fairly. & 1.00 & 18 & 2 & \textbf{11} & 5 \\
Warm / empathetic & Shows empathy, warmth, or emotional support. & .98 & 17 & 6 & 6 & 5 \\
Logical reasoning & Offers logical, rational, or sound reasoning. & .95 & 15 & 5 & 4 & 6 \\
Thoughtful / nuanced & Provides detailed, thoughtful, or nuanced advice. & 1.00 & 13 & 3 & 4 & 6 \\
\bottomrule
\end{tabularx}
\vspace{-0.5em}
\end{table*}

\end{document}